\input harvmac


%
\rightline{EFI-99-8, IASSNS-HEP-99-30}
\Title{
\rightline{hep-th/9903219}
}
{\vbox{\centerline{More Comments on String Theory on $AdS_3$}}}
\medskip

\centerline{\it
David Kutasov${}^1$,
Nathan Seiberg${}^2$}
\bigskip
\centerline{${}^1$Department of Physics, University of Chicago,
5640 S. Ellis Av., Chicago, IL 60637, USA}
\centerline{${}^2$School of Natural Sciences,
Institute for Advanced Study, Olden Lane, Princeton, NJ, USA}
\smallskip

\vglue .3cm
\bigskip
\noindent
We clarify a number of issues regarding the worldsheet and spacetime
descriptions of string propagation on $AdS_3$.  We construct the vertex
operators of spacetime current algebra and spacetime (super) Virasoro
generators in the full interacting $SL(2)$ WZW theory and study their
Ward identities.  We also explain the relation between the analysis in
this note and some recent work on this subject.

\Date{3/99}

\def\journal#1&#2(#3){\unskip, \sl #1\ \bf #2 \rm(19#3) }
\def\andjournal#1&#2(#3){\sl #1~\bf #2 \rm (19#3) }

\def\ie{{\it i.e.}}
\def\eg{{\it e.g.}}

\def\etal{{\it et.al.}}

\def\tilde{\widetilde}

\def\frac#1#2{{#1\over#2}}

\def\half{\frac12}

\def\inbar{\,\vrule height1.5ex width.4pt depth0pt}
\def\IC{\relax\hbox{$\inbar\kern-.3em{\rm C}$}}
\def\IR{\relax{\rm I\kern-.18em R}}
\def\IP{\relax{\rm I\kern-.18em P}}

%
%
\def\np#1#2#3{Nucl. Phys. {\bf B#1} (#2) #3}

\catcode`\@=11
\def\slash#1{\mathord{\mathpalette\c@ncel{#1}}}
\overfullrule=0pt

\def\JJ{{\cal J}}

\def\NN{{\cal N}}

\def\underrel#1\over#2{\mathrel{\mathop{\kern\z@#1}\limits_{#2}}}

\catcode`\@=12


%

\def\exp{{\rm exp}}


\newsec{Introduction}

String theory on $AdS_3$ has many applications.  
It is related to the seminal work of Strominger and Vafa on five
dimensional black holes
\ref\stovafa{A.~Strominger and C.~Vafa, Phys. Lett. {\bf B379}
(1996) 99, hep-th/9601029.}.
It also provides an example of Maldacena's $AdS$/CFT
correspondence 
\ref\maldacena{J.~Maldacena, Adv. Theor. Math. Phys. {\bf 2} (1997)
231, hep-th/9711200.}.
In the context of this correspondence this example is special because
the ``boundary'' (or spacetime) CFT is two dimensional, 
and therefore it has an infinite
dimensional conformal algebra.  This Virasoro algebra was first found
by Brown and Henneaux 
\ref\brownhenneaux{J.D.~Brown and M.~Henneaux,
Commun. Math. Phys. {\bf 104} (1986) 207.},
who studied gravity on $AdS_3$. 
The $AdS_3$ example is furthermore special since it is amenable to a
worldsheet string theory description going beyond the (super) gravity
approximation. The spacetime Virasoro symmetry of string theory on
$AdS_3$ was exhibited in 
\ref\gks{A. Giveon, D. Kutasov, N. Seiberg, 
Adv. Theor. Math. Phys. {\bf 2} (1998) 733, hep-th/9806194.},
where it was also shown that string excitations form representations
of this symmetry, in both bosonic and supersymmetric string theories. 
The work of \gks\ in the bosonic case was later extended in
\ref\dbort{J.~de Boer, H.~Ooguri, H.~Robins and J.~Tannenhauser,
JHEP {\bf 9812} (1998) 26, hep-th/9812046.}. 
The superstring was further discussed by 
\nref\efgt{S. Elitzur, O. Feinerman, A. Giveon and D. Tsabar,
hep-th/9811245.}%
\nref\kll{D. Kutasov, F. Larsen and R. Leigh, hep-th/9812027.}%
\refs{\efgt,\kll} and others. The results of
\gks\ allow one to analyze vacua with Neveu-Schwarz (NS) background fields 
in the NSR formalism.  
Ramond-Ramond (RR) backgrounds were discussed in 
\nref\pes{I. Pesando, hep-th/9809145.}%
\nref\rara{J. Rahmfeld and A. Rajaraman, hep-th/9809164.}%
\nref\pare{J. Park and S. J. Rey, hep-th/9812062.}%
\nref\bvw{N.~Berkovits, C.~Vafa and E.~Witten, hep-th/9902098.}%
\refs{\pes-\bvw}.

One important lesson from these stringy analyses is that (some of) the
observables in the theory correspond to fundamental string excitations
and are thus described by standard vertex operators in the BRST 
cohomology.  Their worldsheet correlation functions, which in flat
backgrounds are usually interpreted as S-matrix elements in target space, 
are interpreted here as Green's functions of operators in the spacetime CFT
\nref\gkp{S.S.~Gubser, I.R.~Klebanov and A.M.~Polyakov,
Phys. Lett. {\bf B428} (1998) 105, hep-th/9802109.}%
\nref\wittenads{E.~Witten, Adv. Theor. Math. Phys. {\bf 2} (1998) 253,
hep-th/9802150.}%
\refs{\gkp,\wittenads}.

In this note we will further extend the work of \gks. We will mainly
use worldsheet techniques, but in this section we start with a target
space analysis, which provides useful guidance to some aspects of
the physics. 

Consider three dimensional gravity with a negative
cosmological constant.  The Lagrangian is
\eqn\minlag{{1 \over \hbar } \sqrt{-g} \big( {2 \over k} + R \big). }
The vacuum solution of its classical equations of motion is $AdS_3$. 
Three dimensional gravity does not have propagating degrees
of freedom.  However, by a careful analysis of the boundary conditions
and the gauge transformations one should mod out by, Brown and Henneaux
showed \brownhenneaux\ that this theory has two Virasoro symmetries
with central charge
\eqn\cent{c = {24\pi\sqrt k \over \hbar}.}
The $\hbar$ in the denominator shows that the effect is classical and
visible at the level of Poisson brackets (as in the computation
of \brownhenneaux).
It is often said that this Virasoro algebra lives at the boundary of
the $AdS$ spacetime.  
More precisely, the statement is that most of the degrees
of freedom of the graviton are pure gauge, with a remnant which can be
gauge transformed to infinity.  In perturbative
string theory, choosing the Lorentz covariant conformal gauge on the
worldsheet leads to Landau gauge in target space.
In this gauge the physical
degrees of freedom of the graviton
are not supported only at the boundary but in the
entire $AdS_3$.

Following Horowitz and Welch 
\ref\horwel{G.T.~Horowitz and D.L.~Welch, Phys. Rev. Lett. {\bf 71}
(1993) 328, hep-th/9302126.},
\minlag\ can be extended (as in string theory) 
by adding the dilaton and
the NS $H=dB$ field
\eqn\stringlag{e^{-2\phi} \sqrt{-g} \big( {4 \over k} + R + 4(\nabla
\phi)^2 - {1 \over 12} H_{\mu\nu\rho}^2 \big), }
where $\hbar$ was absorbed into a shift of $\phi$.  Solving the
equations of motion we find (as usual in string theory) that all the
fields are determined except the zero mode of the dilaton, which is an
integration constant
\eqn\soleq{\eqalign{
&e^{-2\phi} = {1 \over \hbar} \cr
&g_{\mu\nu} =k g_{\mu\nu}^{(0)} \cr
&H ={1 \over \sqrt k} \epsilon = k H^{(0)},\cr}}
where $\epsilon$ is the volume form and $g_{\mu\nu}^{(0)} $ and
$H^{(0)}$ are independent of $k$.  As above, the Brown and Henneaux
analysis leads to a Virasoro algebra with the central charge \cent.   

The spectrum of the theory \stringlag\ includes a single particle, the dilaton,
whose mass is of order $1 \over \sqrt k$.  The dilaton is sometimes
referred to as a fixed scalar.  Despite the fact that the dilaton is
massive, its zero mode $\hbar$ is an arbitrary integration constant.
In other words, the string coupling constant is arbitrary even though
there is no massless field which changes its value\foot{This situation
is similar to the quasi-crystalline compactification to two dimensions
\ref\quasic{J.A.~Harvey, G.~Moore and C.~Vafa, Nucl. Phys. {\bf B304}
(1987) 269.}.}.

The value of $H$ can be interpreted as arising from an electric source
for $H$ of strength
\eqn\defp{p= {4\pi\over \hbar \sqrt{k}}}
with coupling ${p \over 4\pi} \oint B$ at infinity.  Therefore, we can
replace the arbitrary integration constant $\hbar$ by the constant
$p$.  Three dimensional gravity with non-zero $H$ is
analogous to $QED_2$ with a constant background electric field,
$E=\hbar \theta/2\pi$, which can be thought of as resulting from an
electric charge $\theta/2\pi$ at infinity.  $QED_2$ is periodic in
$\theta$; this is a result of spontaneous creation of particles
from the bulk and screening of the source at infinity.
Because of the constant negative curvature in our problem, the volume
is proportional to the area,  and there is no gain in bulk
energy while losing surface energy in the nucleation process.  Hence,
there is no spontaneous nucleation, and we do not expect periodicity
in $p$.  Although spontaneous nucleation does not happen, a long
string can be created with a finite amount of energy
\ref\mms{J. Maldacena, J. Michelson, and A. Strominger,
hep-th/9812073.}.
The nucleation process and the physics of the long string are
discussed in detail in
\ref\seiwit{N. Seiberg and E. Witten, IASSNS-HEP-99-27,  to appear.}.
We will return to the long string in section 7. 

We will see later that, at least in some situations, the integration
constant \defp\ is quantized due to non-perturbative effects.  $p$ is then
interpreted as the number of fundamental strings creating the $AdS$
background. It is natural to conjecture that all such arbitrary
integration constants which are not associated with massless fields
are quantized in string theory.

We now add to \stringlag\ gauge fields of a group $G$,
like those in the NS sector in
string theory.  They are associated with a $\widehat G$ 
current algebra on the worldsheet; the level of $\widehat G$
is an integer $k_G$.  The target space Lagrangian is
\eqn\stringlaggau{e^{-2\phi} \sqrt{-g} \big( {4 \over k} + R +
4(\nabla \phi)^2 - {1\over 12} H_{\mu\nu\rho}^2 + k_G F_{\mu\nu}^2
\big) } 
with
\eqn\modh{H=dB + k_G\omega,}
where $\omega=\Tr (A\wedge dA+{2\over 3}A\wedge A\wedge A)$ 
is the Chern-Simons form of the
gauge fields.  The 
solution of the equations of motion is as in \soleq.  Expanding the
Lagrangian \stringlaggau\ around the classical solution we find the
Chern-Simons interaction
\eqn\csin{{k_Gp\over 4\pi} \omega,}
which leads to masses of order $1\over\sqrt k$ for the gauge fields. 

The Chern-Simons interaction also leads to current algebra on the
boundary
\ref\wittenjones{E.~Witten, Commun. Math. Phys. {\bf 121} (1989) 351.}
with level
\eqn\kst{k_G^{(st)}=k_Gp.}
The result \kst\ was derived in \gks\ using worldsheet techniques.
We now see that the spacetime analysis leads to the same conclusion.

As in the case of the Virasoro algebra discussed above, the spacetime
current algebra can be gauge transformed to the boundary of $AdS_3$. 
In string theory in conformal gauge we are
naturally led to the Landau gauge $\partial ^\mu A_\mu =0$.  The
remaining gauge freedom corresponds to gauge transformations with a harmonic
gauge parameter $\lambda$.  The physical degrees of freedom are
massive spin one particles and various modes which are almost pure
gauge.  These are modes with harmonic $\lambda$ which do not vanish at
infinity.  These modes lead to the current algebra.  
Their profile is non-zero in the bulk of the space-time, as 
we will see below, when we construct the vertex operators of
these modes.

It is now clear that $p$ must be quantized.  This follows either from
unitarity of the spacetime current algebra or from the invariance of
the Lagrangian under large gauge transformations.  The quantization of
$p$ and its interpretation in terms of the number of strings creating
the background are in accord with the general expectation that the
only consistent backgrounds in string theory are those which can occur
due to sources which are themselves dynamical objects in the
theory\foot{Our problem is analogous to the problem 
of the massive IIA theory in 10 dimensions analyzed by
Polchinski and Strominger
\ref\polstrom{J.~Polchinski and A.~Strominger, Phys. Lett. {\bf B388}
(1996) 736, hep-th/9510227.}. They showed that the
source at infinity is quantized in units of the $D8$-branes 
which are present in the theory.}.  
This is another reason to support
the conjecture above that the corresponding 
integration constants in string
theory are quantized. It is also clear that the
quantization of $p$ is non-perturbative in the
coupling, and cannot be seen at any finite order
in the  $\hbar \sim 1/p$ genus expansion of
string theory.

Note that the discussion so far is completely general. It holds for 
any theory of gravity with negative cosmological constant which
is described at low energies by the appropriate Lagrangian, \minlag, 
\stringlag\ or \stringlaggau. Similarly, much of the rest of this note 
is true for any vacuum of string theory that includes an $AdS_3$ factor. 

We now add more structure to \stringlaggau\ by adding to it
supersymmetry.  In particular, we are interested in the
compactification of the type II theory on $T^4\times S^3 $ to three
dimensions.  The bosonic Lagrangian includes the terms \stringlaggau\
for $G=SU(2)_L\times SU(2)_R \times U(1)^4_L \times U(1)^4_R$ gauge
fields coming from Kaluza-Klein reduction of the ten dimensional
metric and NS $B$ field.  
The $SU(2)$ factors arise from the $S^3$ and the $U(1)$
factors from the $T^4$.  The Chern-Simons terms \csin\ for the gauge
groups labeled by $L$ and $R$ have 
opposite signs.  All the massive gauge bosons have mass of order
$1\over \sqrt k$.  Those with one sign of the Chern-Simons term have
spin plus one and the others have spin minus one.
The spacetime theory includes a $\widehat G$ current algebra. The
level \kst\ of the $SU(2)_R$ factor is $k^{(st)}_{SU(2)_R}=kp$ 
($k$ and $p$ are defined as before by \defp, \stringlaggau), while
that of the $U(1)$ factors is $k^{(st)}_{U(1)_R}=p$.  $SU(2)_L$ and
$U(1)_L^4$ have the same levels as $SU(2)_R$ and $U(1)_R^4$ but the
opposite orientations.

Additional gauge fields arise from the RR sector.  These are most easily
analyzed in the IIA language.  In ten dimensions the RR gauge fields are
a one form $A$ and a three form $C$.  The relevant terms in the ten
dimensional Lagrangian are
\eqn\rrterms{{1 \over 2}(dA)^2+ {1 \over 4!} (dC + H \wedge A)^2 +
C\wedge dC \wedge H.} 
For simplicity we consider a square $T^4$ with equal sides $V^{1\over
4}$.  We denote the directions along the $T^4$ by $i,j,k,l$, and the
three non-compact dimensions by $\mu$.  The light
vector fields in three dimensions (which are
$s$-waves on the sphere and constant on the $T^4$)
are: $A= A_\mu$, six fields $A_{\mu,ij}=
C_{ij\mu}$, and $\tilde A_\mu= \epsilon^{ijkl}
\tilde C_{ijkl\mu}$ ($\tilde C$ is a five
form gauge field, which is dual in ten dimensions to $C$).  After
dimensional reduction to three dimensions and a duality transformation
on $C$ we find
\eqn\rrtermst{V (dA)^2+ {1 \over V} (d\tilde A )^2+
(dA_{ij})^2 + k A\wedge d\tilde A + k\epsilon^{ijkl} A_{ij}\wedge
dA_{kl},}
where the terms proportional to $k$ arise from using the expectation
value of $H$ on $S^3$.  

In order to find the spectrum of the theory we should diagonalize the
mass matrix.  We find eight massive spin one particles with mass of
order $1\over \sqrt k$.  Four of them (one linear combination of $A$
and $\tilde A$, and the three $A_{ij}$ which are self-dual on $T^4$)
have spin plus one, and the other four have spin minus one.  We also
find $U(1)_L^4$ and $U(1)_R^4$ current algebras whose level is
$k$, the coefficient of the Chern-Simons terms in \rrtermst.

The full current algebra of the model thus includes a $U(1)_L^8
\times U(1)_R^8$ current algebra, associated with a lattice of
signature $(8,8)$.  In the background we study, which is purely NS,
this lattice is the sum of two signature $(4,4)$ lattices; one for the
NS fields and the other for the RR fields.  The corresponding levels
of the current algebras are $p$ and $k$.

We can also use the above comments to extend the analysis of \kll\
and study the heterotic string on $AdS_3\times S^3\times T^4$.  At
generic points in the Narain moduli space the gauge group is
$SU(2)_L\times SU(2)_R\times U(1)_L^4\times U(1)_R^{20}$, where all
the gauge fields arise from the NS sector.  An analysis of the three
dimensional effective action as above shows that the levels of the two
$SU(2)$ factors are $kp$ and $-(k-2)p$, and the
$U(1)^{24}$ has level $p$ and is related to a signature $(4,20)$
lattice.  This theory is S-dual to the type IIA theory on $
AdS_3\times S^3\times K3$, also with NS background field.  The symmetry is
again $SU(2)_L\times SU(2)_R\times U(1)_L^4\times U(1)_R^{20}$, but
here the $SU(2)_R\times SU(2)_L$ is from the NS sector and has 
level\foot{The difference of $2k$ between the levels of $SU(2)_L$
and $SU(2)_R$ in this case is due to a one string loop correction.
Our formalism should allow one to compute these
corrections by studying the two point functions of various 
currents on the (worldsheet) torus. Since the asymmetry in question
is due to an anomaly, one should compute the contribution of the
odd $(++)$ spin structure to such two point functions. Calculating
this one string loop shift would provide interesting tests of our
formalism and of 
heterotic -- type IIA duality.} $(kp,-k(p-2))$, while the $U(1)_L^4\times
U(1)_R^{20}$ is from the RR sector, still with a signature $(4,20)$
lattice but with level $k$.  These conclusions are consistent with S
duality, which exchanges five branes and one branes, \ie\ $k$ and $p$.

In the rest of this note we will develop a worldsheet description of
string theory on $AdS_3$, which is consistent with the general picture
described in this section and makes it more precise. We will construct
the vertex operators of the affine Lie algebra and Virasoro
generators described above, in the exact worldsheet theory,
and show that their correlation functions satisfy the Ward identities
of two dimensional CFT. 

One important lesson we will learn concerns the origin of the singularities 
in correlation functions in the spacetime CFT.  Such singularites occur when 
two operators approach each other on the boundary.  The locations of the
operators on the boundary are parameters labeling the corresponding vertex 
operators in the first quantized theory. As these parameters are varied,
singularities can occur either because the worldsheet
functional integral over the non-compact target space diverges, or
because the integral over the location of vertex operators on the
worldsheet diverges.  The analysis in \gks\ shows that the
singularities occur from the region where vertex operators approach
each other on the worldsheet.  Thus, short distances on the worldsheet
are mapped by the AdS/CFT correspondence to short distances on the boundary.
The discussion in \dbort\ and its extension below further shows
that the above divergences in the boundary theory
arise because the target space is non-compact and are dominated
by the asymptotic region of the target space. Thus, short distance
behavior on the worldsheet and on the boundary is related to long
distance behavior in the target space gravity theory. 
All this is consistent with -- and nicely demonstrates --
the well known relation between short distance physics on 
the worldsheet and long distance physics in target
space, which is generally true in string theory, and the
UV/IR relation of 
\ref\susswit{L.~Susskind and E.~Witten, hep-th/9805114.},
which relates the short distance regime of the spacetime CFT 
to the asymptotic region of the dual string theory target space.

We start in section 2 by describing classical and quantum CFT on
$AdS_3$.  In section 3 we use this description to construct vertex
operators corresponding to the generators of a spacetime affine Lie
algebra associated with a current algebra on the worldsheet of the
string. In section 4 we prove that these vertex operators satisfy the
correct spacetime current algebra Ward identities and find the
operator that plays the role of the central extension in this current
algebra. In section 5 we study this central extension and compute the
level of the spacetime current algebra.

In section 6 we generalize the construction to the Virasoro case.
We find the vertex operator of the stress tensor of the spacetime 
theory and comment on some of its properties. In section 7 we explain
the relation of the present construction to that of \gks. Section 8 
contains some comments on the generalization of our results to the 
superstring. We summarize our results in section 9.

\newsec{Conformal Field Theory on $AdS_3$} 

\nref\morerefs{
J. Balog, L. O'Raifeartaigh, P. Forgacs, and A. Wipf,
Nucl. Phys. {\bf B325} (1989) 225; 
L. J. Dixon, M. E. Peskin and J. Lykken,  Nucl. Phys. {\bf B325} (1989) 329;
P. M. S. Petropoulos, Phys. Lett. {\bf B236} (1990) 151; 
I. Bars and D. Nemeschansky, Nucl. Phys. {\bf B348} (1991) 89;
S. Hwang, Nucl. Phys. {\bf B354} (1991) 100;  
K. Gawedzki, hep-th/9110076;
J. M. Evans, M. R. Gaberdiel, and M. J. Perry, hep-th/9812252;
I. Bars, Phys. Rev. {\bf D53} (1996) 3308, hep-th/9503205;
in {\it Future Perspectives In String Theory} (Los Angeles, 1995), 
hep-th/9511187.}%

In this section we describe some properties of (worldsheet)
CFT on $AdS_3$, which will be useful later in the
analysis of string theory in this background. We
focus on the Euclidean version of $AdS_3$, the
non-compact manifold $H_3^+= SL(2,C)/SU(2)$. For some early work
on CFT and string theory on $AdS_3$ see \eg\ \morerefs. 

\subsec{ Classical CFT on $AdS_3$}

$H_3^+$ has constant negative curvature $\Lambda=-2/k$ (in
string units). It is described by the metric 
\eqn\metricuy{ds^2=k(d\phi^2 + e^{2\phi} d\gamma d\bar \gamma),}
where $(\phi, \gamma, \bar\gamma)$ are coordinates on
$H_3^+$, with $\phi^*=\phi$, $\gamma^*=\bar\gamma$. 
To define a CFT with this target
space one has to turn on additional $\sigma$-model couplings.
One possibility is to turn on a Neveu-Schwarz $B_{\mu\nu}$ 
field, which leads to the worldsheet Lagrangian
\eqn\convlag{L=k\left(\partial\phi\bar\partial\phi+
e^{2\phi}\bar\partial\gamma\partial\bar\gamma\right).}
The CFT described by \convlag\ has an infinite affine
$SL(2)\times \overline{SL(2)}$ symmetry which, as we will
see below, is very helpful in studying the theory. The global
part of the algebra is generated by
\eqn\jbarj{
\eqalign{
J_0^-&=\partial_\gamma;\;\;J_0^3=\gamma\partial_\gamma
-{1\over2}\partial_\phi;\;\;J_0^+=\gamma^2\partial_\gamma
-\gamma\partial_\phi-e^{-2\phi}\partial_{\bar\gamma}\cr
\bar J_0^-&=\partial_{\bar\gamma};\;\;\bar J_0^3=\bar\gamma
\partial_{\bar\gamma}
-{1\over2}\partial_\phi;\;\;\bar J_0^+=\bar\gamma^2
\partial_{\bar\gamma}
-\bar\gamma\partial_\phi-e^{-2\phi}\partial_\gamma.\cr}}
A large class of observables in this theory is obtained by
studying functions on $H_3^+$. Such functions can be  
decomposed in terms 
of representations of $SL(2)\times \overline{SL(2)}$. 
A convenient tool for performing this decomposition was
proposed in \ref\zamfat{A. B. Zamolodchikov and
V. A. Fateev, Sov. J. Nucl. Phys. {\bf 43} (1986) 657.}
(and further developed in 
\ref\teschner{J. Teschner, hep-th/9712256,  hep-th/9712258;
A. B. Zamolodchikov and Al. B. Zamolodchikov, unpublished.};
see also \ref\morex{J. L. Petersen, J. Rasmussen and M. Yu, 
hep-th/9607129, Nucl.Phys. {\bf B481} (1996) 577; 
O. Andreev, hep-th/9601026, Phys.Lett. {\bf B375} (1996) 60.}).
One introduces an auxiliary complex variable $(x, \bar x)$
and uses the standard representation of the
$SL(2)\times\overline{SL(2)}$ Lie algebra as
the differential operators
\eqn\sltwola{\eqalign{
J_0^-=&-\partial_x;\;\;J_0^3=-(x\partial_x+h);\;\;
J_0^+=-(x^2\partial_x+2hx)\cr
\bar J_0^-=&-\partial_{\bar x};\;\;
\bar J_0^3=-(\bar x\partial_{\bar x}+\bar h);\;\;
\bar J_0^+=-(\bar x^2\partial_{\bar x}+2\bar h\bar x),\cr}}
where $h$ is related to the $SL(2)$ ``spin'' of the 
representation\foot{The value of the quadratic Casimir 
in the representation is $j(j+1)$.} $j$ by $h=j+1$. 
We will refer to
the auxiliary space labeled by $(x,\bar x)$ as
``spacetime'' below, to be distinguished from the
three dimensional target space of the $\sigma$-model,
$H_3^+$, labeled by $(\phi, \gamma, \bar\gamma)$.
$(x,\bar x)$ label the 
space on which the ``dual'' (or spacetime) CFT lives.

One is looking for functions 
$f_{h,\bar h}(\phi,\gamma,\bar\gamma;x,\bar x)$ 
that transform in the spin $(j,\bar j)=(h-1, \bar h-1)$ 
representation of $SL(2)\times \overline{SL(2)}$. The
functions $f_{h, \bar h}$ satisfy the differential 
equations 
\eqn\diffeqn{
\eqalign{
&[J_0^-,f_{h,\bar h}]=-\partial_x f_{h,\bar h} 
= \partial_\gamma f_{h,\bar h}\cr 
&[\bar J_0^-,f_{h,\bar h}]=
-\partial_{\bar x} f_{h,\bar h}
= \partial_{\bar \gamma} f_{h,\bar h}\cr
&[J_0^3\;,f_{h,\bar h}]=-(x \partial_x +h)f_{h,\bar h}
=(\gamma \partial_\gamma-\half\partial_\phi) f_{h,\bar h}\cr 
&[\bar J_0^3\;,f_{h,\bar h}]=-(\bar x \partial_{\bar x} + \bar h)
f_{h,\bar h} 
=(\bar \gamma \partial_{\bar \gamma} - \half \partial_\phi) 
f_{h,\bar h}\cr
&[J_0^+,f_{h,\bar h}]=-(x^2 \partial_x +2h x)f_{h,\bar h}
=(\gamma^2\partial_\gamma  -\gamma 
\partial_\phi  - e^{-2\phi} \partial_{\bar \gamma})f_{h,\bar h}\cr
&[\bar J_0^+,f_{h,\bar h}]=
-(\bar x^2 \partial_{\bar x} +2\bar h \bar x)f_{h,\bar
h}=(\bar \gamma^2\partial_{\bar \gamma} - \bar \gamma \partial_\phi
- e^{-2\phi} \partial_ \gamma) f_{h,\bar h}.\cr}}
The brackets $[\cdots]$ in these equations are
Poisson brackets since we are in the classical theory.
The most general solution of the first 
four equations in \diffeqn\ is
\eqn\firstfour{f_{h,\bar h}
={1 \over (\gamma-x)^h (\bar \gamma - \bar x)^{\bar h}}   
H_{h,\bar h}\left((\gamma-x)(\bar \gamma - \bar x)e^{2\phi}
\right) }
with $H_{h,\bar h}$ an arbitrary function.
The last two equations in \diffeqn\ have solutions {\it only}
for $h=\bar h$, in which case one finds (in a convenient normalization)
\eqn\ooyy{f_{h,h}\equiv
\Phi_h={1\over\pi}
\left({1 \over (\gamma-x)(\bar \gamma -\bar x)e^\phi+e^{-\phi}}
\right)^{2h}.} 
Since $\Phi_h$ is an eigenfunction of the Laplacian in the bulk, it is
the propagator of a particle with mass squared $h(h-1)/k$ 
from the boundary point $x$ to the bulk point
$(\phi,\gamma, \bar \gamma)$.
Note also that \diffeqn\ -- \ooyy\ imply
that one can think of $\Phi_h$ as components of a tensor of weight
$(h, h)$ in spacetime, $\Phi_h(x,\bar x)dx^hd\bar x^h$. This tensor is
non-singular for finite $(x, \bar x)$ (for fixed, finite $(\phi,
\gamma,
\bar\gamma)$). It is also regular at $x\to\infty$; to exhibit
that, redefine $x'=1/x$, and use the transformation
$\Phi'_h(x', \bar x')dx'^hd\bar x'^h=\Phi_h(x, \bar x)dx^hd\bar x^h$.

We can expand $\Phi_h$ around $\phi \approx \infty$
\eqn\limphin{\eqalign{
\Phi_h=&{1\over\pi}\left({1 \over |\gamma-x|^2e^\phi+e^{-\phi}}
\right)^{2h} = \cr 
&{1\over 2h-1}
e^{2(h-1)\phi} \delta^2(\gamma-x) + \CO(e^{2(h-2)\phi}) + {e^{-2h
\phi} \over \pi |\gamma-x|^{4h}} + \CO(e^{-2(h+1) \phi}). }}
For generic $h$ the expansion \limphin\ can be naturally separated
into two independent series. One series includes the leading
term in the $\phi\to\infty$ limit\foot{For $\gamma=x$ the
leading term diverges as $e^{2(h-1)\phi}\delta^2(0)$. 
This reflects the fact that the
dominant term at $\gamma=x$ is $e^{2h\phi}$.  However, since our
vertex operators will always be integrated over, we should use the
expression $e^{2(h-1)\phi} \delta^2(\gamma-x)$ and drop the term
$e^{2h\phi}$.}, $e^{2(h-1)\phi} \delta^2(\gamma-x)$, 
and an infinite series of corrections of the form 
$ e^{2(h-n-1)\phi} \partial_x^n\bar\partial_{\bar x}^n
\delta^2(\gamma-x)$. 
This series is present only in the vicinity of $\gamma=x$. 
The second series starts with the dominant term for generic
$\gamma$, ${e^{-2h \phi} \over |\gamma-x|^{4h}}$, and includes
corrections that are down by $e^{-2n\phi}$ with $n\in Z_+$.

The situation here is similar to that in Liouville
theory 
\ref\mssm{N. Seiberg, Prog. Theor. Phys. Supp. {\bf 102} (1990) 319; 
G. Moore, N. Seiberg and M. Staudacher, \np{362}{1991}{665}.}.
The analogs of the wavefunctions $\Phi_h$ \ooyy\ in that case
are the exact minisuperspace wavefunctions, the modified
Bessel functions $K_\nu$. 
The operators \ooyy\ with real $h$ are the analogs
of microscopic operators in Liouville theory. 
The analog of the expansion \limphin\ is the expression
of $K_\nu$ as a linear combination of $I_{\pm\nu}$. 
As in \limphin, for generic $\nu$ expanding $I_{\pm\nu}$
around $\phi\to\infty$ gives rise to two infinite series that
do not mix. For integer $\nu$ the two series mix and
interesting ``resonances'' occur. Analogous phenomena
exist in our case; the two series in \limphin\ mix
when $h\in Z/2$. 

As we approach $h=1/2$ in \ooyy, the behavior of $\Phi_h$
changes:
\item{(1)} For $h>\half$ (and $\gamma\simeq x$)
the leading term at large $\phi$ is
$e^{2(h-1)\phi} \delta^2(\gamma-x)$. 
\item{(2)} For $h=\half$ we have a ``resonance'' 
($e^{2(h-1)\phi}=e^{-2h\phi}$) and the leading term
in the expansion of \limphin\ is $2\phi e^{-\phi} \delta^2(\gamma-x)$.
This is consistent with the fact that 
the coefficient of $e^{2(h-1)\phi}\delta^2(\gamma-x)$ in \limphin\
diverges as $h\to\half$.
\item{(3)} For $0<h<\half$ the leading term is ${e^{-2h \phi}
\over |\gamma-x|^{4h}} $.  However, the integral 
$\int d^2\gamma\Phi_h$ over $|\gamma-x|<\epsilon$ 
does not vanish as $\epsilon\to 0$. Therefore, one
cannot neglect the contribution
${1\over 2h-1} e^{2(h-1)\phi} \delta^2(\gamma-x)$
to $\Phi_h$. 
\item{(4)} For $h=0$ we have a more dramatic ``resonance'' and
$\Phi_{0}=1$. 
\item{(5)} For $h<0$ the leading term at large $\phi$ 
and generic $\gamma$,
${e^{-2h \phi} \over |\gamma-x|^{4h}}$, diverges as $\phi\to\infty$.  
\item{(6)} For all values of $h$ discussed above, 
$\Phi_h e^\phi$ grows as $\phi\to\infty$. Thus $\Phi_h$ 
is a non-normalizable operator, as is familiar from 
Liouville theory and the AdS/CFT correspondence. 

Since 
\sltwola\  $J_0^-=-{\partial\over \partial x}$, 
$\bar J_0^-=-{\partial\over \partial \bar x}$,
every observable 
$\Theta(x, \bar x)$ is conjugate to $\Theta(0)$
\eqn\xevol{\Theta(x, \bar x)=e^{-xJ_0^--\bar x \bar J_0^-} 
\Theta(0) e^{xJ_0^-+\bar x\bar J_0^-}.}
Therefore, it is natural to define
\eqn\jofxdef{\eqalign{ J^+(x;z)=&e^{-xJ_0^-} J^+(z) e^{xJ_0^-}=J^+(z)
-2xJ^3(z)+x^2J^-(z)\cr 
J^3(x;z)=&e^{-xJ_0^-} J^3(z) e^{xJ_0^-}=J^3(z) -xJ^-(z) = 
-\half \partial_x J^+(x;z)\cr 
J^-(x;z)=&e^{-xJ_0^-} J^-(z) e^{xJ_0^-}=J^-(z)
=\half \partial_x^2 J^+(x;z).\cr}}
Since all these currents are related to each other by differentiation, 
it is enough to consider
\eqn\intoppp{J( x; z)\equiv -J^+(x; z)=2 x J^3(z)- J^+(z) - x^2 J^-(z).}
An explicit computation shows that
\eqn\intopppe{J( x; z)=k\left[(x-\gamma)^2e^{2\phi}\partial_z\bar \gamma
+2(x-\gamma)\partial_z \phi-\partial_z\gamma\right].}
Similar expressions can be written for $\bar J(\bar x;\bar z)$.

$J(x;z)$ is a spacetime tensor of weight $(-1,0)$ 
which, just like $\Phi_h$, is
non-singular for all $x$. 
The transformation properties \diffeqn\ of $\Phi_h$ under 
$SL(2)\times \overline{SL(2)}$ can be summarized by the following 
action of the charges $J_0(x)$ associated with the conserved 
currents \intoppp
\eqn\sumtrans{\left[J_0(x), \Phi_h(y,\bar y)\right]=\left[
(y-x)^2\partial_y+2h(y-x)\right]\Phi_h(y,\bar y)}
and a similar relation for $\bar J_0$. 

An observable which will be important below is 
\eqn\jphizdef{\bar J(\bar x;\bar z)\Phi_1(x,\bar x; z,\bar z).}
It is easy to check that
\eqn\JPhiLambda{{\pi\over k}\bar J\Phi_1= \partial_{\bar z} \Lambda,}
where
\eqn\ddmm{\Lambda = -{1 \over \gamma-x}
{(\gamma-x)(\bar \gamma - \bar x)e^{2\phi} \over (\gamma-x)(\bar
\gamma - \bar x)e^{2\phi}+1}.}
Since $\Phi_1$ is a tensor of weight $(1,1)$ in spacetime,
while $\bar J$ is a tensor of weight $(0, -1)$, \JPhiLambda\
suggests that $\Lambda$ is a tensor of weight $(1,0)$. However, 
the operators \ooyy\ are all left-right symmetric and  as indicated
above there are no solutions of \diffeqn\ for $h \not=\bar h$.  What
is then the status of $\Lambda$ of \ddmm?
One can check that $\Lambda$ transforms as follows under 
$SL(2)\times\overline{SL(2)}$
\eqn\classsltwo{\eqalign{
\left[J_0(x),\Lambda(y, \bar y)\right]=&\left[(y-x)^2\partial_y+2(y-x)
\right]\Lambda(y, \bar y)-1\cr
\left[\bar J_0(\bar x),\Lambda(y, \bar y)\right]=
&(\bar y-\bar x)^2\partial_{\bar y}
\Lambda(y, \bar y).\cr}}
Comparing to \sumtrans\ we see that $\Lambda$ indeed transforms 
like an object with $\bar h=0$ under $\overline{SL(2)}$, while its
transformation as an object with $h=1$ 
under $SL(2)$ contains an anomalous term, which is
a constant 
(independent of the worldsheet fields and of $x$, $y$). 
Thus, it is natural to expect that $\Lambda$ is {\it not} a
good observable in CFT on $AdS_3$. Later we will provide two 
independent arguments that support this conclusion. However,
we will also see that the object of interest to us will be
$\partial_{\bar z}\Lambda$ which, as is clear from 
\JPhiLambda, {\it is}
a good observable in the theory\foot{As a good analogy consider 
the operators $X$ and $\partial_{\bar z}X$ in two dimensional
massless scalar field theory. While $X$ is not a good
observable, \eg\ because its two point function is logarithmic
in the separation, $\langle X(z) X(w)\rangle\sim \log|z-w|^2$,
$\partial_{\bar z}X$ is a good observable.}. This is consistent with the
fact that the anomalous contribution to $[J_0^+,\Lambda]$,          
being constant, disappears when we differentiate with respect to
$\bar z$ as in \JPhiLambda. We also note for later use that:
\item{(1)} As we approach the boundary of $AdS_3$,
\eqn\ddmml{\lim_{\phi\rightarrow \infty} \Lambda = {1 \over
x-\gamma}.} 
\item{(2)} As functions on $AdS_3$,
$\Lambda$ and $\Phi_1$ \ooyy\ are related by
\eqn\ddll{\partial_{\bar x}\Lambda=\pi\Phi_1.} 

\subsec{ Quantum CFT on $AdS_3$}

In the quantum theory, 
the currents \jbarj\ are 
generators of an $\widehat{SL(2)}\times \widehat{\overline{SL(2)}}$ 
algebra. Their OPE's are\foot{Here and below we denote by $\sim$
equality up to non-singular terms in an OPE.} 
\eqn\opealg{
\eqalign{
J^3(z) J^\pm(w)\sim &{\pm J^\pm(w)\over z-w} \cr
J^3(z) J^3(w)\sim &-{{k\over2}\over (z-w)^2}\cr
J^-(z)J^+(w)\sim &{k\over (z-w)^2}+{2J^3(w)\over z-w}.\cr}}
The operators $\Phi_h$ are 
primaries of the full $\widehat{SL(2)}\times 
\widehat{\overline{SL(2)}}$ algebra\foot{The operator that
is usually referred to as a primary of $\widehat{SL(2)}$ is
$\Phi_h(x=0)$. $\Phi_h(x)$ is related to it by conjugation
\xevol.}. They satisfy
\eqn\rrr{\eqalign{
J^3(z) \Phi_h(x,\bar x;w,\bar w)\sim &-{(x\partial_x+h)\Phi_h(x,\bar x)
\over z-w}\cr
J^+(z) \Phi_h(x,\bar x;w,\bar w)\sim &-{\left(x^2\partial_x+2hx\right)
\Phi_h(x,\bar x)\over z-w}\cr
J^-(z) \Phi_h(x,\bar x;w,\bar w)
\sim &-{\partial_x\Phi_h(x,\bar x)\over z-w}\cr
}}
and similar OPE's with the $\bar J^A$. All operators in CFT
on $AdS_3$ can be obtained by acting on the primaries $\Phi_h$
by the currents $J(x;z)$, \intoppp, and $\bar J(\bar x; \bar z)$.

The stress tensor of the worldsheet CFT is given by the
Sugawara form
\eqn\stresst{T^{\rm ws}(z)={1\over k-2}\eta_{AB}J^AJ^B=
{1\over k-2}\left[ -(J^3)^2
+J^+J^-\right].}
$T^{\rm ws}$ satisfies the Virasoro algebra with $c=3k/(k-2)$.
There is
a similar expression for $\overline T^{\rm ws}$ in terms of $\bar J$.
The worldsheet scaling dimensions of $\Phi_h$, 
$(\Delta_h, \bar\Delta_h)$ can be computed using
\stresst. They are
\eqn\one{\Delta_h=\bar\Delta_h=-{h(h-1)\over k-2}.}

The operator product expansions \opealg\ and \rrr\ can be expressed in
a more compact form using the operator $J$ of \intoppp
\eqn\JJ{J( x; z) J( y; w)\sim k {(y-x)^2\over( z- w)^2}+ { 1\over
z- w}\left[(y-x)^2\partial_y -2(y-x) \right] J( y; w)}
\eqn\JPhi{J( x; z) \Phi_h(y,\bar y;w,\bar w)\sim  {1 \over z-w} \left[(
y-x)^2\partial_y +2h(y-x)\right]\Phi_h(y, \bar y) .}
These expressions are in accord with the interpretation of $J(x)$ as
$J^+$ conjugated to the point $x$ \jofxdef.  
This point of view also makes it obvious
why the OPE's \JJ, \JPhi\ are regular when $x=y$.

The stress tensor \stresst\ can be written
directly in terms of the
current $J(x;z)$ as
\eqn\newstresst{T^{\rm ws}={1\over 2(k-2)}
\left[J \partial_x^2 J -\half (\partial_x J)^2\right].}
A few comments:
\item{(1)} Using $\partial_x^3J=0$, $\partial_x T^{\rm ws}=0$.
The combination \newstresst\ corresponds to a tensor in spacetime
of weight $(0,0)$. 
\item{(2)}
The composite operator $ J(
x)\Phi_h(x,\bar x)$ does not require normal ordering,
and the classical manipulations in
\jphizdef\ and \JPhiLambda\ are justified in the quantum theory.
Thus \JPhiLambda\ holds in the quantum theory (without any finite
renormalizations). A related fact is that $J( x)\Phi_h(x,\bar 
x)$ is a primary of worldsheet Virasoro \stresst. 
\item{(3)} Useful special cases of \JJ, \JPhi\ are
\eqn\usespec{\eqalign{
J(x;z)\Phi_1(y,\bar y; w,\bar w)&\sim 
{1\over z-w}\partial_y\left[(x-y)^2\Phi_1(y,\bar y;w,\bar w)\right]\cr
J(x;z)\left[J(y;w)\Phi_1(y,\bar y; w,\bar w)\right]&\sim 
k(x-y)^2\partial_w\left[{\Phi_1(y, \bar y; w, \bar w)\over
z-w}\right].\cr }}
\item{(4)} In the free field Wakimoto representation
\ref\wakim{M. Wakimoto, Comm. Math. Phys. {\bf 104} (1986) 605.}, 
\intoppp\ takes the form ($\alpha_+=\sqrt{2k-4}$)
\eqn\wakimform{J(x;z)=-\beta(x-\gamma)^2+\alpha_+\partial_z\phi
(x-\gamma)-k\partial_z\gamma,}
which is related to the classical expression \intopppe\ by the
appropriate rescaling of the variables, certain finite
renormalizations, and the equations of motion of the Wakimoto
representation, which imply 
$\beta=-ke^{2\phi\over\alpha_+}\partial_z\bar\gamma$.

\noindent
The two point function of $\Phi_h$ is
determined by worldsheet conformal invariance and
the $SL(2)\times \overline{SL(2)}$ action \rrr.
It is
\eqn\twoptfn{\langle \Phi_h(x,\bar x;z,\bar z)
\Phi_{h'}(y,\bar y;w,\bar w)\rangle=\delta(h-h')
{D(h;k)\over |x-y|^{4h}|z-w|^{4\Delta_h}}.}
The numerical coefficient $D(h;k)$ has been computed
in \teschner.  Note that for $h=h'$ the two point function is
infinite -- it is proportional to $\delta(0)$.   This infinite
factor can be interpreted as the volume of the subgroup of
the $SL(2,C)$ isometry of our target space,
which preserves $x$ and $y$.  

When considering string theory on $AdS_3$ one has to integrate over
the locations of all vertex operators on the worldsheet and to divide
by the volume of a subgroup of the two dimensional diffeomorphism
group isomorphic to the Mobius group $SL(2,C)_M$.  For the zero point
function, the CFT correlation function is proportional to the volume
of the target space; \ie\ to the volume of $SL(2,C)$.  Dividing by the
volume of $SL(2,C)_M$ we find a finite answer.  Similarly, a one
point function, if it does not vanish, is proportional to the subgroup
of $SL(2,C)$ which leaves one point in the target space invariant.
Dividing by the volume of the subgroup of $SL(2,C)_M$ which leaves a
point on the worldsheet invariant we again find a finite answer.
Finally, for the two point function, the CFT correlation function is
proportional to $\delta(0)$, as above, but in string theory this
infinity is cancelled by a similar factor from $SL(2,C)_M$.  All this
is exactly as in non-critical string theory \mssm.

One can also compute the OPE of $J$ with $\Lambda$ by integrating
\usespec\ with respect to $\bar y$. The $\bar y$ independent
integration constant can be determined by comparing with the
semiclassical analysis \classsltwo, or by using the free field
representation \wakimform. This leads to
\eqn\JLambda{J( x; z) \Lambda(y,\bar y;w,\bar w)\sim  {\left[( y-
x)^2\partial_y+2(y-x)\right]\Lambda(y, \bar y;w,\bar w)-1\over z-w} .}
We can now provide an independent argument that
$\Lambda$ itself is not a good observable in the theory.
According to \JPhiLambda, \ddll, \twoptfn\ the two point
function of $\Lambda$ satisfies 
\eqn\twoptlam{\langle\Lambda(x;z)\Lambda(y;w)\rangle
={a\log|z-w|^2+b\log|x-y|^2\over (x-y)^2},}
where $a$ and $b$ are infinite constants that become
finite when passing to string theory, as discussed
above. We have also used the fact that correlation functions
are single valued both on the worldsheet and in spacetime.  
The correlator \twoptlam\ depends logarithmically on the 
separation, both on the worldsheet and in spacetime. Hence, like
$X$ in free field theory, $\Lambda$ is not a good observable.

Another argument that $\Lambda$ cannot be a good observable
in CFT on $AdS_3$ follows from the relation \ddll.
$\Phi_1$ is used in string theory on $AdS_3$ to dress marginal
operators in the spacetime CFT. If $\Lambda$ was a good
observable, all dimension $(1,1)$ operators in the
spacetime CFT would have been total derivatives with respect to
$\bar x$ of operators with scaling dimensions $(1,0)$. 
This is highly unlikely based on the available information
about particular examples of string theory on $AdS_3$ (see
\eg\ \gks\ for details). If one further assumes that  
the spacetime CFT is unitary, any operator of dimension
$(1,0)$ must be holomorphic (in $x$), and therefore we 
would have concluded 
that there are {\it no} marginal operators at all in the spacetime
CFT, which again contradicts available information.

We next turn to the OPE algebra of the $\widehat{SL(2)}$ primaries 
$\Phi_h$. De Boer \etal\ \dbort\ studied the operator product 
$\Phi_{h_1}(x_1,\bar x_1;z_1,\bar z_1)\Phi_{h_2}(x_2,\bar x_2; z_2,\bar z_2)$ 
in the semiclassical region of large $k$, for $h_2$ of order $k$ and $h_1$ 
of order one. The insertion of $\Phi_{h_2}$ affects the worldsheet and takes 
$\phi(z_2)\rightarrow \infty$ and $\gamma(z_2)\rightarrow x_2$.  
Therefore, the operator product expansion 
is dominated by large $\phi$ and $\gamma\approx x_2$,
\eqn\debb{\Phi_{h_1}(x_1;z_1)
\Phi_{h_2}(x_2;z_2) \sim |z_1-z_2|^{-4{(h_1-1)(h_2-1)\over k-2}}
\delta^2(x_1-x_2) \Phi_{h_1+h_2-1}(x_2;z_2),} 
where we inserted the factor
$|z_1-z_2|^{-4{(h_1-1)(h_2-1)\over k-2}}$ on dimensional grounds.
We can also study this operator product when $h_1,h_2 \ll k$ in the
semiclassical region.  In this limit the dimensions of all $\Phi_h$
are much smaller than one.  Therefore, the leading contribution comes 
{}from terms with no derivatives in the operator product expansion
$\Phi_{h_1} \Phi_{h_2}$.  Terms with derivatives have larger dimensions.  
At large $\phi$, the product behaves like
\eqn\prodV{\eqalign{
\Phi_{h_1} \Phi_{h_2}=&
e^{2(h_1+h_2-2)\phi} \delta^2(\gamma-x_1) \delta^2(\gamma-x_2)
+\CO(e^{2(h_1+h_2-3)\phi})\cr
=& e^{2(h_1+h_2-2)\phi} \delta^2(x_1-x_2)
\delta^2(\gamma-x_2) +\CO(e^{2(h_1+h_2-3)\phi}).\cr}}  
Therefore, it is
approximately $\Phi_{h_1}(x_1;z_2)\Phi_{h_2}(x_2;z_2) =
\delta^2(x_1-x_2) \Phi_{h_1+h_2-1} + ...$ where the ellipses stand for
operators with $\Re h<h_1+h_2-1$.  When $k$ is large but not infinite
the dimensions of the operators do not vanish and the right hand side
should be multiplied by $|z_1-z_2|^{-4{(h_1-1)(h_2-1) \over k-2}}$,
as in \debb.  The
operators with $\Re h<h_1+h_2-1$ have larger dimension than
$\Phi_{h_1+h_2-1}$ and therefore they are negligible as $z_1\rightarrow
z_2$. We conclude that at least for $h_1\ll k$ and $k\gg 1$
\eqn\opesem{\eqalign{
&\Phi_{h_1}(x_1, \bar x_1;z_1, \bar z_1) 
\Phi_{h_2}(x_2,\bar x_2; z_2, \bar z_2) =\cr
&C_{12}|z_1-z_2|^{-4{(h_1-1)(h_2-1)\over k-2}} \delta^2(x_1-x_2)
\Phi_{h_1+h_2-1}(x_2,\bar x_2; z_2, \bar z_2) 
+ {\rm smaller~ terms ~as~} z_1\rightarrow z_2.\cr}} 
For small values of $h$ the asymptotic form of the vertex operators is
different (see the discussion above) and this conclusion could be
modified. 
A special case of \opesem\ is
\eqn\opesems{\lim_{z_1\rightarrow z_2} 
\Phi_1(x_1,\bar x_1; z_1, \bar z_1) \Phi_h(x_2,\bar x_2; z_2, \bar
z_2) = \delta^2(x_1-x_2) \Phi_h(x_2,\bar x_2; z_2, \bar z_2).}
An important aspect of \opesems\ is that the 
coefficient on the right hand side is independent of $h$ and therefore
we can normalize $\Phi_1$, as in \limphin, such that it is equal to one.

In the exact theory the situation is more complicated. The OPE
structure is \teschner
\eqn\opeexact{\eqalign{
\Phi_{h_1}(x_1, \bar x_1;z_1, \bar z_1)  &
\Phi_{h_2}(x_2,\bar x_2; z_2, \bar z_2) = \cr
& \int dh_3\int d^2x_3 {f_{h_1h_2h_3}(x_i, \bar x_i)\Phi_{h_3}(x_3,
\bar x_3;z_2, \bar z_2) \over
|z_1-z_2|^{2(\Delta_1+\Delta_2-\Delta_3)}} + {\rm descendants}.}}
The OPE coefficients $f_{h_1h_2h_3}(x_i, \bar x_i)$ can be computed
by studying the three point functions of $\Phi_{h_j}$ in CFT
on $AdS_3$. They have two components. The one
that is studied in \teschner\ corresponds to $f_{123}$ which is a smooth
function of $x_i$ and of $h_i$. Later we will be interested in
computing the OPE \opesems\ in the exact theory.
Smooth $f$'s do not contribute to this quantity since
one could continue analytically in $h$ from a region
where the limit of \opesems\ as $z_1\to z_2$ vanishes. 

A second class of $f$'s corresponds to distributions in
$x_i$. In particular, we can examine the possibility
that  
\eqn\distrib{\Phi_{h_1}(x_1, \bar x_1;z_1, \bar z_1) 
\Phi_{h_2}(x_2,\bar x_2; z_2, \bar z_2) \sim 
{f(x_1, x_2) \over |z_1-z_2|^{2(\Delta_1+\Delta_2-\Delta_3)}}
\Phi_{h_3}(x_2, \bar x_2;z_2, \bar z_2).}
In that case, $SL(2)$ invariance constrains severely 
the function $f(x_1, x_2)$ and \opesem\ is one of the
only consistent solutions of the $SL(2)$ Ward identity.
Indeed, applying $\oint J(y)$ to the two sides of the OPE 
\distrib\ where the contour         
surrounds $z_1$, $z_2$, we find
that $f$ must satisfy
\eqn\wico{[ (x_1-y)^2\partial_{x_1}+(x_2-y)^2\partial_{x_2}
+2h_1(x_1-y)+2h_2(x_2-y)-2h_3(x_2-y) ] f=0.}
The coefficient of $y^2$ in \wico\ (the $J^-$ Ward identity) 
shows that $f$ depends only on $x_1-x_2$.  The coefficient of 
$y$ in \wico\ (the $J^3$ Ward identity) leads to
\eqn\wicoo{[(x_1-x_2) \partial_{x_1} +(h_1+h_2-h_3)]f=0.}
Therefore, if $f$ is not a distribution,
\eqn\simplepower{f=(x_1-x_2)^{h_3-h_1-h_2}.}
The term without $y$ in \wico\ is then satisfied only if
$h_2=h_1+h_3$. For the particular case of interest in \opesems,
$h_1=1$, we find that $h_2=h_3+1$, and the power of $|z_1-z_2|$
in \distrib\ is $4h_3/(k-2)>0$. Thus, such an $f$ does not contribute
to the limit \opesems.  

If we also allow $f$ which is a distribution, \wicoo\ can be satisfied
by
\eqn\stisb{f=\partial_{x_1}^n\partial_{\bar x_1}^{\bar n}
\delta^2(x_1-x_2) \qquad {\rm for}~ h_1+h_2-h_3=n+1.}
A similar relation with $\bar J$ sets $n=\bar n$.  The term with no
$y$ in \wico\ further allows the answer to be of this form only for
$n=0$, $h_1+h_2-h_3=1$ or $h_1=(n+1)/2$, $h_3=h_2-(n+1)/2$.

To summarize, we see that terms of the form \distrib\ in the OPE
$\Phi_1\Phi_h$ give a finite contribution in the limit $z_1\to
z_2$ only if $x_1$ is in the vicinity of $x_2$. The corresponding
OPE coefficient is proportional to $\delta^2(x_1-x_2)$, as in 
\opesems. Note that it is not guaranteed a-priori that one
can choose the numerical prefactor on the right hand side 
of \opesems\ to be one (it could depend on $h$). 
We saw that this prefactor can be chosen to be one in the
semiclassical limit, and we will assume that this continues to hold
in the exact theory. This assumption is necessary for the spacetime
Ward identities.

The above discussion also shows that the divergence of the
OPE coefficient in \opesems, $\delta^2(x_1-x_2)$, as 
$x_1\to x_2$ is directly related to the non-compactness
of $AdS_3$. Short distance singularities in the spacetime
CFT come from $\phi\to\infty$ on $AdS_3$. As discussed in the
introduction, this is an example of the UV/IR correspondence
of \susswit.

Another description of CFT on $AdS_3$ is obtained by 
``Fourier transforming'' the local fields $\Phi_h$ and defining
the mode operators 
\eqn\qqq{V_{j;m,\bar m}=\int d^2x
x^{j+m}\bar x^{j+\bar m}\Phi_{j+1}(x,\bar x).}
The inverse transformation is
\eqn\ppp{\Phi_h(x,\bar x)=\sum_{m,\bar m}V_{h-1;m,\bar m} x^{-m-h}
\bar x^{-\bar m-h}.} 
Equations \qqq, \ppp\ are essentially mode expansions of the 
local spacetime fields in the dual CFT and, just as in standard
CFT, the local fields are the fundamental objects and the
modes may or may not be well defined and useful. Indeed, in \ppp\ it is
in general not clear what range the variables $(m,\bar m)$ should run
over\foot{Locality in $(x,\bar x)$ space implies that
$m-\bar m\in Z$, but $m+\bar m$ is not constrained.}. 
In \qqq\ the integral over $x$ is in general divergent and is 
defined by finding a region where it converges and analytically
continuing from there. This may or may not be possible in different cases.
Thus, we will here work
with local fields in $x$ space and not with their modes.
The modes $V_{j;m,\bar m}$ have the scaling dimensions \one\
with $h=j+1$.
They satisfy the OPE algebra with the currents (which is equivalent
to \rrr, \JPhi)
\eqn\onea{\eqalign{
J^3(z) V_{j;m,\bar m}(w,\bar w)\sim &{m V_{j;m,\bar m}\over z-w}\cr
J^\pm(z) V_{j;m,\bar m}(w\bar w)\sim &{(m\mp j)V_{j;m\pm 1,\bar m}\over
z-w}\cr }}
and similarly for $\bar J^A(\bar z)$. In addition to primaries, 
the theory contains descendants which
have higher order poles in their OPE with $J^A$. The latter
can be described as (normal ordered) products of 
currents and their derivatives with $V_{j;m,\bar m}$. 

\newsec{The spacetime current algebra} 

We next turn to (bosonic) string theory on
$AdS_3\times \NN$, with $\NN$ an arbitrary compact
manifold which together with $AdS_3$ provides a solution
of the string equations of motion (see \gks\ for a more
detailed discussion). Assume that the CFT on $\NN$ 
contains a worldsheet current algebra associated with some group 
$G$, generated by worldsheet currents $k^a(z)$ satisfying
\eqn\wsope{k^a(z)k^b(w)\sim {\half k_G\delta^{ab}\over (z-w)^2}
+{f^{abc} k^c(w)\over z-w},}
where $k_G$ is the level of the worldsheet affine Lie algebra 
$\widehat G$ and ${1\over i}f^{abc}$ are the (real)
structure constants of $G$. One can then construct in the spacetime theory 
currents $K^a(x)$, which satisfy an algebra similar to \wsope.
The purpose of this section is to construct the generators of this 
spacetime current algebra, following \gks. On general grounds one expects 
the local field $K^a(x)$ to be described by the vertex operator 
of an almost pure gauge field. We start with a review of some relevant 
properties of gauge fields in string theory.

Associated with each worldsheet current $k^a(z)$ satisfying \wsope\
one finds in string theory a target space gauge field whose vertex 
operator is
\eqn\gengauge{\int d^2z A^a_\mu(x)k^a(z)\partial_{\bar z} x^\mu,}
where $x^\mu$ are coordinates on the target space\foot{In our
case the target space is $AdS_3$, and $(x^\mu)= 
(\phi,\gamma, \bar\gamma)$.}.
Under infinitesimal gauge transformations the gauge field
$A^a_\mu$ transforms as\foot{For simplicity, we neglect the non-linear term
in the gauge transformation, \ie\ assume that
$A$ is small.}
\eqn\gaugetr{\delta A^a_\mu=\partial_\mu\lambda^a.}
Thus, a pure gauge field \gengauge\ is described by the vertex
operator 
\eqn\puregauge{\int d^2z k^a(z)\partial_{\bar z}\lambda^a(x^\mu),}
where $\lambda^a$ is the gauge function. The fact that the
theory is gauge invariant implies that couplings like
\puregauge\ do not influence the physics, however this
is only true for gauge functions $\lambda^a$ with compact support
in target space. As we will see in detail
below, gauge transformations with gauge function $\lambda^a$ that
do not vanish at the boundary of $AdS_3$ generate
a large algebra of global transformations, the affine Lie algebra
$\widehat G$.
 
There is an important point that should be mentioned here.
Eqs. \gengauge, \puregauge\ describe gauge field couplings in
the $\sigma$-model. In
string theory, one has to impose the worldsheet consistency conditions.
In particular, the integrands of \gengauge, \puregauge\ must be
primaries under the left and right-moving worldsheet Virasoro 
symmetries
with scaling dimension $(1,1)$.  When the target space is flat, this
implies that the gauge field satisfies the gauge condition
$\partial^\mu A_\mu^a=0$ (as well as the massless Klein-Gordon
equation $\partial_\nu\partial^\nu A_\mu^a=0$). 
The remaining gauge freedom corresponds to gauge functions satisfying 
$\partial_\mu\partial^\mu\lambda^a =0$. If the gauge function
$\lambda^a$ has compact support, it can be Fourier transformed. The
corresponding operators are $\int d^2z k^a(z) \partial_{\bar z}
e^{ik\cdot x}$ with $k^2=0$.  Moding out by BRST commutators removes
such operators.  

Another set of ``pure gauge'' operators satisfying
$\partial_\mu\partial^\mu\lambda^a=0$ is associated
with $\lambda^a=x^\mu$ and $\lambda^a=1$.  
These gauge functions do not go to zero at infinity. The 
corresponding vertex operators, $\int d^2 z k^a(z)
\partial_{\bar z} x^\mu$ and $\oint dz k^a(z)$, are in the BRST
cohomology.  
The status of the above two kinds of operators 
in string theory is actually 
somewhat different. The gauge functions $\lambda^a=x^\mu$
are not good operators in the worldsheet conformal field theory.
This can be seen by noticing that their two point functions depend on
the logarithm of the separation.  Therefore, the 
corresponding ``pure gauge'' vertex operators
are not BRST exact and give rise to physical vertex
operators (the zero momentum gluons). The gauge function
$\lambda^a=1$ does correspond to a good operator on the worldsheet 
(the identity operator), and the corresponding vertex operator
$\int d^2zk^a(z)\partial_{\bar z}1$ vanishes. Its vanishing
is the statement that all correlation functions in string theory
are invariant under the global part of the gauge symmetry. 

Returning to string theory on $AdS_3$,
we expect the spacetime affine currents to have the form
\eqn\genkb{K^a(x)=
\int d^2 zk^a(z)\partial_{\bar z}
\lambda(\gamma,\bar\gamma,\phi; x, \bar x) }
with a gauge function $\lambda$ that does not go to zero as
$\phi\to\infty$.  Since $K^a(x)$ has spacetime scaling dimension
$h=1$, $\bar h=0$, we are looking for an operator $\partial_{\bar
z}\lambda$ with worldsheet scaling dimension
$(\Delta,\bar\Delta)=(0,1)$ and spacetime scaling dimension $(h, \bar
h)=(1,0)$. We constructed precisely such an operator in \JPhiLambda,
\ddmm. Below we will determine the normalization $\lambda=-{1\over
\pi} \Lambda$,
\eqn\genkcc{K^a(x)=-{1\over \pi}
\int d^2 zk^a(z)\partial_{\bar z}
\Lambda(\gamma,\bar\gamma,\phi; x, \bar x).}
The spacetime
current \genkcc\ is formally pure gauge, but the
gauge function $\Lambda$ does not have compact support, and 
is not a good observable in CFT on $AdS_3$. The corresponding 
vertex operator $K^a(x)$ does not decouple from correlation functions.  
Another useful way of writing $K^a(x)$ (using \JPhiLambda) is
\eqn\ddllff{K^a(x)=-{1\over k}\int d^2z k^a(z) \bar J(\bar x; \bar z)
\Phi_1(x,\bar x; z, \bar z),}
which makes manifest the fact that $K^a(x)$ is physical and
has weight $(1,0)$ in spacetime. 

{}From the point of view of the spacetime theory we expect
the current $K^a(x)$ \genkcc\ to be holomorphic, up to contact
terms. Differentiating
\genkcc\ with respect to $\bar x$ and using \ddll\ gives
\eqn\genkeee{\partial_{\bar x} K^a=-
\int d^2z k^a(z)\partial_{\bar z}\Phi_1.}
Unlike \genkcc, $\partial_{\bar x} K^a$ given by \genkeee\
is a total derivative of a good observable in CFT on $AdS_3$,
and we expect it to decouple (for generic $x$). Indeed we will
see in the next section that \genkeee\ vanishes inside correlation
functions, except for contact term contributions at the locations
of other operators in the correlation function. These contact terms
give rise to the affine Lie algebra Ward identities familiar from
two dimensional CFT.

To summarize, ``pure gauge'' vertex operators
\puregauge\ in string theory belong to one of a few classes, 
depending on properties of the gauge function $\lambda(x)$.
If $\lambda$ is normalizable, \puregauge\ is null and decouples.
This is the situation usually considered in string theory.
If $\lambda$ is non-normalizable but is a good observable
in the worldsheet CFT, like $\Phi_1$ \genkeee,
the pure gauge field \puregauge\
decouples ``generically,'' \eg\ for generic momenta or
for generic $x$ in the $AdS_3$ example \genkeee.
If the gauge function $\lambda$ is not a good observable
in the worldsheet theory, but $\partial_{\bar z}\lambda$
is (as in \genkcc, \ddllff), 
the gauge field \puregauge\ does not decouple and
is in the BRST cohomology. Gauge functions $\lambda$
for which $\partial_{\bar z}\lambda$ is not a good
observable (\eg\ $\lambda=x^\mu x^\nu$ in flat spacetime)
should not be considered, since in that case \puregauge\
is not a physical operator.

\newsec{The current algebra Ward identity}

Now that we have constructed the spacetime affine Lie algebra 
generator $K^a(x)$ we would like to study its correlation
functions. The typical correlation function of interest is
\eqn\wardid{\langle K^a(x) K^b(y) W_h(s, \bar s)\cdots \rangle,}
where $W_h(s, \bar s;w, \bar w)$ is an operator that transforms in
some representation $R$ of the worldsheet current algebra,
\eqn\repW{k^a(z)W_h(s, \bar s; w, \bar w)\sim 
{t^a(R)W_h(s, \bar s; w, \bar w)\over z-w}.}
$t^a(R)$ are the representation matrices of the Lie algebra
of $G$ in the representation $R$.
$h$ labels the transformation properties of $W_h$ under
the worldsheet $\widehat{SL(2)}\times\widehat{\overline{SL(2)}}$
algebra; in other words\foot{For simplicity, we will restrict
to $W_h$'s that are primary under worldsheet $\widehat{SL(2)}
\times \widehat{\overline{SL(2)}}$. As explained in \gks, there
are many physical operators that are descendants of worldsheet
current algebra but are physical in string theory (\ie\ primary
under worldsheet Virasoro).}, $W_h(s)$ includes a factor of $\Phi_h(s)$.
General considerations in the spacetime CFT imply that the
correlation function \wardid\ should satisfy the standard
current algebra Ward identities, 
\eqn\stkmwi{\eqalign{
K^a(x)K^b(y)\sim &{\half k_G^{(st)}\delta^{ab}\over (x-y)^2}
+{f^{abc} K^c(y)\over x-y}\cr
K^a(x)W_h(s, \bar s)\sim &          
{t^a(R)W_h(s, \bar s)\over x-s}.\cr }}
Our main task in this section is to derive this Ward
identity in string theory on $AdS_3$, and to compute
the level of the spacetime current algebra
$k_G^{(st)}$.

The strategy of the calculation is the following. The 
integral in \genkcc\ runs over the whole worldsheet, 
$\Sigma$. We would like
to write $K^a(x)$ as an integral of a total derivative. 
Since $\partial_{\bar z}k^a(z)=0$ except for delta function
contributions at the locations on the worldsheet of the other
insertions in the correlator, we can write the
integrand of \genkcc\ as a total derivative if we restrict 
the worldsheet integral to $\Sigma'$, the Riemann
surface $\Sigma$ with small holes around the insertions removed.
The integral \genkcc\  
over such holes goes to zero as their size 
is taken to zero; therefore, extracting the holes does not
change the value of correlators like \wardid\ and amounts
to a particular regularization. 

After extracting the holes we can rewrite \genkcc\ as
\eqn\genkc{K^a(x)=-{1\over\pi}\int_{\Sigma'} d^2z
\partial_{\bar z}\left[
k^a(z)\Lambda(x,\bar x;z,\bar z)\right].}
If the worldsheet correlators of $\Lambda$ are well behaved,
we can perform the integral in \genkc\ and it will receive 
contributions only from the boundaries of the Riemann surface
with holes $\Sigma'$. This leads to the representation
\eqn\genkd{K^a(x)=\sum_i\oint_{C_i} {dz\over2\pi i} k^a(z)\Lambda(
x,\bar x;z,\bar z),}
where $C_i$ is the boundary of the small hole around the $i$'th
insertion. 

Previously we have argued that $\Lambda$ is
not a good observable in CFT on $AdS_3$; \eg\ it has
a logarithmically divergent two point function \twoptlam.
However, the question that is relevant for analyzing the validity
of \genkd\ is how $\Lambda(z,\bar z)$ behaves as one approaches
an insertion involving $\Phi_h$ or some 
(worldsheet) current algebra descendant
thereof. To see that near such insertions $\Lambda$
is well behaved, one can integrate \opesems\ with respect to
$\bar x_1$, using
\eqn\deltaf{\partial_{\bar x}{1\over x-y}=\pi\delta^2(x-y),}
and fix the $\bar x_1$ independent integration
constant by a combination of semiclassical and free field
techniques. This leads to
\eqn\opeLPhi{\lim_{z_1\rightarrow z_2} 
\Lambda(x_1,\bar x_1; z_1, \bar z_1) \Phi_h(x_2,\bar x_2; z_2, \bar z_2)
\sim  {1\over x_1-x_2} \Phi_h(x_2,\bar x_2; z_2, \bar z_2).}
Alternatively, one can differentiate \genkd\ with respect to $\bar x$
and use \ddll. One finds
\eqn\genke{\partial_{\bar x}
K^a(x)=\sum_i\oint_{C_i} {dz\over2i} k^a(z)\Phi_1(
x,\bar x;z,\bar z),}
which is equivalent to \genkeee. 
Here there are no subtleties regarding the existence of the 
right hand side in CFT on $AdS_3$.
The two representations \genkd, \genke\ are of course equivalent
but below we will find it convenient to use \genkd\
in the asymptotic analysis near $\phi=\infty$ 
(in section 7) and \genke\ in the
discussion of the exact theory (in this section).

Note that \genke\ and the related form \genkeee\ 
provide an example of the worldsheet -- spacetime
relation discussed in \gks. The left hand side of \genkeee\ is naively
zero since $K^a(x)$ is a holomorphic current in spacetime.
Correlation functions of $\partial_{\bar x}K^a$ such as \wardid\
receive contact term contributions from the boundaries of spacetime,
which correspond to $x$ approaching other insertions. These contributions
are governed by the spacetime Ward identities.
The right hand side of \genkeee\ is naively zero since 
$k^a(z)$ is a holomorphic current\foot{Or, equivalently, 
the right hand side of \genkeee\ is BRST exact on the worldsheet.}. 
Correlation functions of the right hand side 
of \genkeee\ receive contributions from the boundaries of the
worldsheet, which correspond to $z$ approaching other
insertions. These contributions are governed by the worldsheet
Ward identity.
 
In the rest of this section we will derive the spacetime 
Ward identity for removing $K^a(x)$ from correlation functions
like \wardid\ by using the representation \genke\ for $K^a$.
Substituting $\partial_{\bar x}
K^a(x)$ \genke\ in \wardid\ we have to analyze the contributions 
to the 
correlator from the boundaries of the worldsheet, $C_i$. Consider 
the contribution from a small circle around the location of 
$W_h(s,\bar s)$
\eqn\smallcontW{\langle\partial_{\bar x} K^a(x) W_h(s)\cdots\rangle
\sim \int d^2w\oint_w{dz\over 2i}\langle k^a(z)\Phi_1(x,\bar x;z, \bar
z) W_h(s,\bar s; w, \bar w)\cdots\rangle.}
Using \opesems, \repW\ and performing the integral over $z$ leads
to
\eqn\opeKW{\langle \partial_{\bar x} K^a(x) W_h(s)\cdots\rangle
\sim \pi\delta^2(x-s) t^a(R)\langle W_h(s)\cdots\rangle,}
which is the correct result \stkmwi\ (using \deltaf).

The contribution to the correlator \wardid\ from a small circle
surrounding the location of another current, $K^b(y)$, can
be computed similarly
\eqn\opeKK{\langle\partial_{\bar x} K^a(x) K^b(y)\cdots\rangle
\sim -{1\over k}\int 
d^2w\oint_w{dz\over 2i}\langle k^a(z)\Phi_1(x,\bar x;z, \bar z)
k^b(w)\bar J(\bar y; \bar w)\Phi_1(y,\bar y; w, \bar w)\cdots\rangle.}
There are now two contributions, from the single and double pole
in the worldsheet OPE \wsope. The single pole is treated in a 
similar way to the one discussed for $W_h$ above. The only
new element here is that there might be 
an additional contribution from the OPE of $\bar J(\bar y; \bar w)$
with $\Phi_1(x,\bar x;z, \bar z)$. Since that OPE has a positive 
power of $x-y$ and it multiplies a $\delta^2(x-y)$ \opesems, this contribution
vanishes. Thus in direct analogy to \opeKW\ one finds that the single
pole contribution gives
\eqn\opeKKsingle{\langle \partial_{\bar x} K^a(x) K^b(y)\cdots\rangle
=\pi\delta^2(x-y) f^{abc}\langle K^b(y)\cdots\rangle+\cdots .}
The double pole contribution to \opeKK\ gives an operator proportional to 
\eqn\newstr{\lim_{z\to w} \partial_z\Phi_1(x,\bar x;z,\bar z)
\bar J(\bar y;\bar w)\Phi_1(y,\bar y;w,\bar w),}
which we would like to compute using \opesems. Unfortunately,  
the limit \newstr\ is sensitive to the $O(z-w)$ term in \opesems\
which of course is suppressed as $z\to w$.  
Nevertheless, we can make use of \opesems, by 
using the fact that
\eqn\parphi{\partial_z\Phi_1(x,\bar x;z, \bar z)
={1\over k}\partial_x\left[ J(x;z)\Phi_1(x,\bar x;
z,\bar z)\right].}
This can be shown by computing the OPE of 
the worldsheet stress tensor \stresst\ 
with $\Phi_1$ using \rrr\ and isolating the single pole term,
which by definition is $\partial_z\Phi_1$. 
An alternative proof of \parphi\ is obtained by
using the fact that $\pi\Phi_1=\partial_x\bar\Lambda$ (the
complex conjugate of \ddll); thus 
\eqn\derivpart{\partial_z\Phi_1={1\over\pi}\partial_x\partial_z\bar
\Lambda={1\over k}\partial_x(J\Phi_1),}
where in the last step we used the complex conjugate of 
\JPhiLambda.
Subsituting this form of $\partial_z\Phi_1$ in \newstr\ 
and using \opesems\ we find that
\eqn\newstraa{\eqalign{
&k\lim_{z\to w} \partial_z\Phi_1(x,\bar x;z,\bar z)
\bar J(\bar y;\bar w)\Phi_1(y,\bar y;w,\bar w)\cr
&=\lim_{z\to w}\partial_x\left[ J(x;z)\Phi_1(x,\bar x;                 
z,\bar z)\bar J(\bar y;\bar w)\Phi_1(y,\bar y;w,\bar w)\right]\cr
&=\partial_x\delta^2(x-y)J(y;w)\bar J(\bar y;\bar w)
\Phi_1(y,\bar y; w, \bar w).\cr }} 
Finally, collecting all the terms and coefficients, we find
that the spacetime currents satisfy 
\eqn\opeKKfull{\langle \partial_{\bar x} K^a(x) K^b(y)\cdots\rangle
\sim \pi\delta^2(x-y) f^{abc}\langle K^b(y)\cdots\rangle-
\pi\half k_G\delta^{ab}\partial_x\delta^2(x-y)\langle I
\cdots\rangle,} 
where the operator $I$ is given by
\eqn\newp{I={1\over k^2}
\int d^2z J(x;z)\bar J(\bar x;\bar z)\Phi_1(x,\bar x;
z,\bar z).}
Thus, the currents $K^a(x)$ satisfy the spacetime OPE algebra
\eqn\spkakb{K^a(x)K^b(y)\sim 
{f^{abc}\over x-y} K^c(y)+{\half Ik_G\delta^{ab}\over
(x-y)^2}.}  
In the next section we 
discuss the operator
$I$ \newp\ that appears in the spacetime current algebra,
and study the consequences of \spkakb\ for correlation functions
in string theory on $AdS_3$. 

\newsec{Properties of the central extension}

The discussion of the previous sections makes it
clear that the operator $I$ \newp\ has the following properties:
\item{(1)} It is physical (BRST invariant) in string theory on
$AdS_3$.  Naively $I$ is BRST exact because of \JPhiLambda, but like 
the spacetime current $K^a(x)$ it does not decouple, for the same
reasons. 
\item{(2)}  As is clear from the form of the vertex operator, $I$ is
essentially a mode of the dilaton/graviton field.
\item{(3)} It transforms as a tensor of weight $(0,0)$ in spacetime,
\ie\ its scaling dimensions in the spacetime CFT are $h=\bar h=0$.
Therefore, we intepret it as (a multiple of)
the identity operator of the spacetime
CFT. 

\noindent
For this interpretation to be consistent, $I$ must be a
constant, $\partial_x I=\partial_{\bar x}I=0$. To prove that $I$ is
indeed independent of $x$, $\bar x$, 
one can use considerations similar to those of the
previous sections. By using
\JPhiLambda\ we can write $\partial_{\bar x}I$ (up to a constant,
which is irrelevant for the argument) as
\eqn\nmwp{\partial_{\bar x}I\propto\int d^2z J(x;z)\partial_{\bar
z}\Phi_1 \propto\int d^2z \partial_{\bar
z}\left[J(x;z)\Phi_1\right]\propto\oint dz \partial_z\bar\Lambda.}
So, $\partial_{\bar x} I$ is given by an integral
of a total derivative over the boundary of the worldsheet. 
This will vanish, unless there is a (worldsheet) singularity in the OPE 
of the operator $\bar\Lambda$ with other
physical operators in the theory. But we have used
above the fact that such OPE's are in fact regular
(see \eg\ \opeLPhi). Hence, the operator
\nmwp\ actually decouples in all correlation functions 
and we conclude that $\partial_{\bar x} I=0$. A similar
argument can be used to show that $\partial_x I=0$. 

The fact that $I$ is a constant means that it is not a vertex operator
of a scalar particle in $AdS$.  Above we noticed that it is a mode of
the spacetime dilaton field.  These two assertions are consistent
because the dilaton is a fixed scalar and hence it is massive.  As a
massive particle it does not have an on-shell zero mode.  $I$ is a
physical on-shell zero mode of the dilaton field.  It is related to the
integration constant discussed in section 1.  Below we will write the
vertex operator of the massive dilaton.

This similarity and distinction between the massive dilaton and $I$ is
identical to the relation between the massive gauge bosons and the
current algebra vertex operators $K^a(x)$.  The latter are vertex
operators of modes of the gauge fields, which should not be
confused with the vertex operators of the massive gauge fields.  Below
we will write the vertex operators of the massive gauge fields.

Could the constant $I$ be zero? The two point function
$\langle I(x) I(y)\rangle$ is proportional to 
$\langle\Phi_1(x,\bar x; z, \bar z)\Phi_1(y, \bar y; w, \bar w)\rangle$
which cannot  
vanish\foot{Vertex operators of marginal operators in the
spacetime CFT are products of marginal operators on the worldsheet
and $\Phi_1$.
Their two point functions
lead to the Zamolodchikov
metric and are proportional to $\langle
\Phi_1\Phi_1\rangle$. Thus the latter cannot vanish.}. 
Therefore, $I$ is a non-zero multiple
(which will be computed below) of the identity operator of the
spacetime theory.

To see the consequences of the OPE \spkakb\ for correlation
functions in string theory on $AdS_3$, consider a correlation
function of the form
\eqn\gencor{G^{ab}(x,y, x_1,\cdots, x_n)\equiv
\langle\langle K^a(x) K^b(y) V_{h_1}(x_1)\cdots V_{h_n}(x_n)\rangle
\rangle,}
where $V_{h_j}(x_j)$ are any physical operators in string theory on
$AdS_3$, and the double bracket notation implies that \gencor\
receives contributions from all worldsheet topologies including
disconnected ones (this point was
emphasized in \dbort). As $x\to y$, $G^{ab}$ exhibits the following
behavior
\eqn\gencora{\eqalign{
&G^{ab}(x,y, x_1,\cdots, x_n)=\cr
&{\half k_G\delta^{ab}\over (x-y)^2} 
\langle\langle IV_{h_1}(x_1)\cdots V_{h_n}(x_n)\rangle
\rangle+{f^{abc}\over x-y}
\langle\langle K^c(y) V_{h_1}(x_1)\cdots V_{h_n}(x_n)\rangle
\rangle+{\rm finite}.\cr}}
Thus, the level of the spacetime current algebra $k_G^{(st)}$
is
\eqn\stlevel{k_G^{(st)}=k_GP(g_s),}
where $P(g_s)$ is defined by
\eqn\iconst{\langle\langle IV_{h_1}(x_1)\cdots V_{h_n}(x_n)\rangle
\rangle=P(g_s)\langle\langle V_{h_1}(x_1)\cdots V_{h_n}(x_n)\rangle
\rangle.}
This is the statement that $I$ is a 
multiple of the identity operator.   In order to explore \iconst\ and
to express $P(g_s)$ in terms of correlation functions of $I$, we
consider for simplicity a case where the
disconnected correlation functions of the $V_{h_i}$
all vanish.  It is simple to generalize the discussion to the case
where they do not vanish.   Then,
\eqn\maniiconst{\langle\langle IV_{h_1}(x_1)\cdots V_{h_n}(x_n)\rangle
\rangle=\langle\langle I\rangle\rangle\langle\langle
V_{h_1}(x_1)\cdots V_{h_n}(x_n)\rangle\rangle +\langle\langle I
V_{h_1}(x_1)\cdots V_{h_n}(x_n)\rangle\rangle_{\rm connected},}
where the last term is a sum over all genera but only of connected
worldsheets. Comparing \iconst\ an \maniiconst\ we see that connected
correlation functions must satisfy 
\eqn\assumei{\langle\langle I V_{h_1}(x_1)\cdots V_{h_n}(x_n)
\rangle\rangle_{\rm connected} =I(g_s)
\langle\langle V_{h_1}(x_1)\cdots V_{h_n}(x_n)
\rangle\rangle_{\rm connected}}
with $I(g_s)$ independent of the operators $V_{h_i}$.
This is a strong assumption, which we do not know how to prove, but
below we will give some evidence that it is true.  
Equations \iconst, \maniiconst\ and \assumei\ imply that
\eqn\inconstbe{P(g_s)= \langle\langle I \rangle\rangle + I(g_s).}  
To leading order in $g_s$ (or equivalently $1/p$ \defp), 
$P(g_s)=\langle I\rangle_0=I_0Z_0$, where $I_0$ is the 
normalized one point function of $I$ on the sphere, 
\eqn\iooo{I_0\equiv{\langle I\rangle_0\over Z_0},}
where $Z_0$ is the spherical partition sum.
Similarly, the leading order form of \assumei\ is
\eqn\assumesphere{\langle I V_{h_1}(x_1)\cdots V_{h_n}(x_n)
\rangle_0 =I_0 \langle V_{h_1}(x_1)\cdots V_{h_n}(x_n)\rangle_0}
{}from which we see that the leading behavior of $I(g_s)$
as $g_s\to 0$ is $I(g_s)=I_0+O(g_s^2)$. The leading
term is fixed by setting all $V_h$ to one in \assumesphere\
and comparing to \iooo.

The assumption \assumei\ can be tested and the value of $I_0$ can be
determined by comparing different correlation functions. Consider for
example the following three point functions\foot{The
spherical correlation functions in the
next few equations are divided by the partition
sum $Z_0$, so that $\langle 1\rangle_0=1$.} and use \assumesphere
\eqn\thpta{\eqalign{
\langle K^{a_1}(x_1) K^{a_2}(x_2)K^{a_3}(x_3)\rangle_0
=&{\half k_GI_0 f^{a_1a_2a_3}\over (x_1-x_2)(x_1-x_3)(x_2-x_3)}\cr 
\langle K^{a_1}(x_1) K^{a_2}(x_2)I\rangle_0 =&{\half k_GI_0^2 
\delta^{a_1a_2}\over (x_1-x_2)^2}\cr 
\langle III\rangle_0=&I_0^3 .\cr}} 
A direct calculation of the correlators \thpta\ using
worldsheet techniques gives
\eqn\thptb{\eqalign{
\langle K^{a_1}(x_1) K^{a_2}(x_2)K^{a_3}(x_3)\rangle_0
=&{\half k_G {C_{111}\over k^2}
f^{a_1a_2a_3}\over (x_1-x_2)(x_1-x_3)(x_2-x_3)}\cr
\langle K^{a_1}(x_1) K^{a_2}(x_2)I\rangle_0
=&{\half k_G {C_{111}\over k^3}\delta^{a_1a_2}\over (x_1-x_2)^2}\cr
\langle III\rangle_0=&{C_{111}\over k^4}, \cr}}
where $C_{111}$ is defined by
\eqn\fofofo{\langle\Phi_1(x_1, \bar x_1;z_1,\bar z_1)
\Phi_1(x_2, \bar x_2; z_2, \bar z_2)
\Phi_1(x_3, \bar x_3; z_3, \bar z_3)\rangle_0={C_{111}\over |x_1-x_2|^2|x_1-x_3|^2
|x_2-x_3|^2}.}
Comparing \thpta\ and \thptb\ we find three equations
for the two unknowns $C_{111}$ and
$I_0$. The fact that a solution exists,
\eqn\thptd{C_{111}=k;\;\;\; I_0={1\over k},}
provides a consistency check of the assumption \assumesphere.

The leading order contribution to the spacetime
level \stlevel\ is thus $k_GZ_0I_0 $.  
The partition sum on the sphere can presumably be calculated
using worldsheet methods; this has not been done to date and we
will not attempt to compute it here. Alternatively, one can use the
fact that the partition sum of any string theory on the sphere
is equal to the classical action evaluated on the solution
of the equations of motion. Substituting \soleq\ into \stringlag\
gives zero, but the action has a finite 
surface term of order $kp$.  Therefore, 
$Z_0\simeq kp$ up to a numerical coefficient (independent
of $k$, $p$) which we will not attempt to fix. This leads to 
\eqn\finksp{k_G^{(st)}\approx k_G Z_0I_0=k_Gp}
(where the $\approx$ sign refers to the large $p$ approximation) 
in agreement with the spacetime result \kst. The numerical
coefficient, which has not been computed, can be fixed either by a
calculation of the worldsheet partition sum or by a Gibbons-Hawking
type calculation of the classical action corresponding to
\stringlag, \soleq; the latter is known to give the correct answer
\finksp.

\newsec{The spacetime Virasoro symmetry}

In this section we generalize the results of the previous
sections to the case of the spacetime Virasoro symmetry.  
The discussion is analogous to the one for affine Lie
algebras, so we mention only the new aspects. 
Using the expected spacetime scaling 
dimension of the spacetime stress tensor
$T(x)$ we can write it
as a linear combination
\eqn\virasa{T(x)=\int d^2z \left[A_1
J(x;z)\partial_x^2\Phi_1+A_2\partial_xJ\partial_x\Phi_1+
A_3\partial_x^2J\Phi_1\right] \bar J(\bar x;\bar z).}
BRST invariance imposes a constraint on the coefficients $A_i$.  
One solution, with the overall normalization fixed below, is
\eqn\viras{\eqalign {
T(x)=&{1\over2k}\int d^2z \left[
J(x;z)\partial_x^2\Phi_1+3\partial_xJ\partial_x\Phi_1+
3\partial_x^2J\Phi_1 \right] \bar J(\bar x;\bar z) = \cr
&{1\over2}\oint {dz\over 2\pi i}\left[
J(x;z)\partial_x^2\Lambda+3\partial_xJ\partial_x\Lambda+
3\Lambda\partial_x^2J \right].}}
We have already seen above that $\partial_x\int d^2z J\bar J\Phi_1 =
\int d^2z\partial_z (\bar J \Phi_1)=0$.  Therefore, we have the
freedom to shift 
$T$ by any constant times $\int d^2z \partial_x^2(J \Phi_1)\bar J$
(this operator is BRST exact and vanishes in correlation functions).  
Using this freedom we can write
\eqn\virasn{T(x)={1\over2k}\int d^2z
\left(\partial_xJ\partial_x\Phi_1+ 2\partial_x^2J\Phi_1 \right) \bar
J(\bar x;\bar z) = {1\over2}\oint {dz\over 2\pi
i}\left(\partial_xJ\partial_x\Lambda+  2\Lambda\partial_x^2J \right).}
Note that: 
\item{(1)} $T(x)$ is a tensor of weight $(2,0)$ in spacetime. In fact,
imposing that it should be such a tensor is an alternative
way of fixing the relative coefficients in \virasa. 
\item{(2)} One might wonder whether it is possible to generate
tensors of weight $(h,0)$ with $h>2$ in string theory on $AdS_3$ 
by generalizing the logic of \virasa\ -- \virasn. It is easy to see 
that there are no new operators (in addition to $\partial_x^n T(x)$)
at higher levels; the reason is that $\partial_x^3 J(x;z)=0$.

\noindent
It is usually said that in string theory on $AdS_3$ (with $B_{NS}$
turned on) the dilaton is a fixed scalar, meaning that it is
massive. These statements refer {\it not} to the vertex operators
\newp,\virasn, which are almost pure gauge. The vertex
operator of the massive dilaton is
\eqn\massdil{D(x,\bar x)=\int d^2z
\left(\partial_xJ\partial_x+ 2\partial_x^2J\right)
\left(\partial_{\bar x}\bar J\partial_{\bar x}+ 
2\partial_{\bar x}^2\bar J\right)\Phi_1,}
which has spacetime scaling dimension $(2,2)$ 
and is only quasi-primary under the spacetime Virasoro
algebra -- it has the quantum numbers of
$T(x)\bar T(\bar x)$. 
A similar statement can be made regarding the gauge bosons
associated with a worldsheet affine Lie algebra $k^a(z)$.
The gluons get a mass from the Chern-Simons term discussed
in section 1. The corresponding vertex operators are
\eqn\massgauge{A^a(x,\bar x)=\int d^2z k^a(z)
\left(\partial_{\bar x}\bar J\partial_{\bar x}\Phi_1+ 
2\partial_{\bar x}^2\bar J\Phi_1\right).}
The operators $A^a$ have scaling
dimension $(1,2)$ in the spacetime CFT
and have the quantum numbers of $K^a(x)\bar T(\bar x)$.

However, $D(x,\bar x)\not=T(x)\bar T(\bar x)$ because the
two point function of $D(x,\bar x)$
scales like $1/g_s^2$, while that of $T\bar T$ arises from a disconnected
diagram and scales like
$1/g_s^4$. Similarly,  $A^a(x,\bar x)\not=
K^a(x)\bar T(\bar x)$.

To derive the Virasoro Ward identity it is convenient to insert
into correlators the operator $\partial_{\bar x} T$,
\eqn\revisdd{\partial_{\bar x}T(x)={1\over2}\oint {dz\over 2i}\left[
\partial_xJ\partial_x\Phi_1+
2\Phi_1\partial_x^2J
\right].}
A calculation of the sort outlined above for the affine Lie algebra 
case can now be performed for the stress tensor and one finds the
correct Ward identities for $TT$ and $TW_h$, 
\eqn\stvirwi{\eqalign{
T(x)T(y)\sim&{\half c^{(st)}\over (x-y)^4}
+{2T(y)\over (x-y)^2}+{\partial_y T\over x-y}\cr
T(x)W_h(s, \bar s)\sim&          
{hW_h(s, \bar s)\over (x-s)^2}+{\partial_s W_h\over x-s}\cr
}}
with $c^{(st)}=6kP(g_s)$.

\newsec{Relation to \gks}

The original analysis of \gks\ is valid in the free field region of
$\phi \rightarrow \infty$.  Therefore, in order to compare our exact
discussion above with \gks\ we should first consider the behavior of 
all our operators in that limit. 

Using $\lim_{\phi \rightarrow \infty}\Lambda = 1/(x-\gamma)$ \ddmml\
in the expression for the current \genkcc\ we find
\eqn\newll{K_f^a(x)\equiv \lim_{\phi \rightarrow \infty}
K^a(x)=-{1\over\pi}\int_{\Sigma'} d^2z k^a(z) \partial_{\bar z}{1\over
x-\gamma(z)},} 
or after performing the integration by parts \genkc, \genkd
\eqn\nmm{K_f^a(x)=\sum_i\oint_{C_i}{dz\over 2\pi i}{k^a(z)\over
x-\gamma(z)}.}
Here we can see the relation to \gks\ by expanding \nmm\ in powers of
$x\over \gamma$ or $\gamma \over x$; \ie\ performing the Fourier
transform discussed in section 2, to find
\eqn\kmmgks{K^a_n = \oint {dz\over 2\pi i}k^a(z) \gamma^n.}
Recalling that $\gamma(z)$ is holomorphic in the $\phi\to\infty$
limit, we can rewrite \newll\ as
\eqn\mmn{K_f^a(x)=\int_{\Sigma'} d^2z 
k^a(z) \partial_{\bar z}\bar\gamma\delta^2(x-\gamma(z)).}
Using
\eqn\useli{\lim_{\phi\rightarrow \infty} \Phi_1 = \delta^2(\gamma-x)}
and the expression for $\bar J$, \intopppe, we
recognize \mmn\ as the limit as $\phi\to\infty$ of \ddllff.

Similarly, using \useli\ and \intopppe\ in the exact expression for $I$
\newp\ we find
\eqn\centch{I_f\equiv
\lim_{\phi\rightarrow \infty} I = \int_{\Sigma'}  d^2z
\delta^2(x-\gamma) \partial \gamma \bar \partial \bar \gamma =
-{1\over \pi} \int_{\Sigma'}  d^2z
\partial_z \gamma  \partial_{\bar z}{1 \over x-\gamma(z)} =
\sum_i\oint_{C_i} {dz\over 2\pi i}{\partial_z \gamma \over
x-\gamma(z)}.}
Expanding this in powers of $x\over \gamma$ we find the Fourier modes
\eqn\Ine{I_n= \oint {\partial_z \gamma \over \gamma} \left({x\over
\gamma}\right)^n = \delta_{n,0}\oint {\partial_z \gamma \over
\gamma}} 
in agreement with the corresponding expression in \gks. 

The analysis of the Ward identities of the previous sections
simplifies considerably in the $\phi\rightarrow \infty$ limit and
becomes identical to \gks\ (up to the Fourier transform from
$x$ space to the modes). Indeed, consider the $\phi \rightarrow 
\infty$ limit of the correlation function \wardid
\eqn\wardidf{\langle K_f^a(x) K_f^b(y) W_h(s, \bar s)\cdots
\rangle}
with operators $W_h(s)$, which are proportional to
$e^{2(h-1)\phi}\delta^2(\gamma-x)$ (see, \limphin).  (In \gks\ the
Fourier modes of these operators \qqq\ 
were used.)  The contribution to \wardidf\
{}from a contour $C$ around one of the $W_h$ operators is 
\eqn\kaw{K_f^a(x)W_h(s)\sim 
\int d^2w\oint_w {dz\over 2\pi i} {k^a(z)\over x-\gamma(z)} 
W_h(s, \bar s;w, \bar w) ={t^a(R)\over x-y}\int d^2w W_h(s, \bar s;w,
\bar w),}  
where we used the OPE \repW\ and the fact that $W_h$ is proportional
to a delta function.  We also used the fact that there is no short
distance singularity in the OPE $\gamma(z) \gamma(w)$ (again in the
free field region $\phi\to\infty$).

A similar calculation of the contribution from the region where
$K_f^a$ approaches on the worldsheet another current
$K_f^b$ is most conveniently done by using the representation
\nmm\ for $K_f^a$ and \mmn\ for $K_f^b$. There are now two
contributions coming from the single and double poles in the
worldsheet OPE \wsope.  They lead to
\eqn\kakb{\eqalign{
K_f^a(x)K_f^b(y)\sim &
\int d^2w\oint_w {dz\over 2\pi i} {k^a(z)\over x-\gamma(z)}
k^b(w)\partial_{\bar w} \bar\gamma\delta^2(y-\gamma(w))\cr
=& {f^{abc}\over x-y}\int d^2w k^c(w)\partial_{\bar w}
\bar \gamma \delta^2(y-\gamma(w))+{\half
k_G\delta^{ab}I_f\over(x-y)^2}. \cr}}

Thus, \kaw, \kakb\ lead to the spacetime Ward identities
\eqn\sptiwi{\eqalign{
K_f^a(x)W_h(s)\sim &{t^a(R)\over x-y} W_h(s)\cr
K_f^a(x)K_f^b(y)\sim &{f^{abc}\over x-y} K_f^c(y)+{\half I_f
k_G\delta^{ab}\over (x-y)^2},\cr}}
which are the $\phi\to\infty$ analogs of \stkmwi, \spkakb.
The Virasoro generator $T(x)$ constructed in section 6
also simplifies in the limit $\phi\to\infty$. Using \ddmml\ in
\virasn\ one finds
\eqn\txf{T(x)=\half \oint {dz\over 2\pi i}
\left[-{\partial_x J\over (x-\gamma)^2}+{2\partial_x^2J\over
x-\gamma}\right].}
Since \intoppp\ $\partial_x J=2J^3-2xJ^-$, $\partial_x^2J=-2J^-$,
expanding in powers of $\gamma/x$ leads to
\eqn\txfgks{-L_n=\oint {dz\over 2\pi i}\left[
(n+1)J^3\gamma^n-nJ^-\gamma^{n+1}\right]}
in agreement with \gks. The OPE algebra \stvirwi\ can be 
derived similarly to the affine Lie algebra case discussed 
above; as there, the derivation is related to the discussion 
of \gks\ by a Fourier transform.

So far we have shown how the exact expressions in the previous sections
simplify as $\phi \rightarrow \infty$ and reproduce the (Fourier
transform to $x$ space of the) corresponding expressions in \gks.
We next look more closely at the relation between the two calculations.

The vertex operators in the theory can be determined in the free field
region $\phi \rightarrow \infty$. 
The quantum numbers of these vertex operators can
also be determined in the asymptotic region.  However, their
correlation functions depend on the entire space and cannot be
computed using only the free field region.  An exception to this is
Ward identities like \sptiwi.  These are associated with
singularities in $x$ space.  As explained in the introduction, such
singularities are determined by the $\phi \rightarrow \infty$ limit.
For holomorphic operators like $K^a(x)$ the correlation functions are
completely determined by the singularities and therefore the
asymptotic analysis suffices.  The rest of the information in the
correlation functions depends on the full structure of the theory
including the bulk of the target space where $\phi$ is small.
We conclude that the asymptotic expressions of the vertex
operators, as in \gks, are good for some purposes like enumerating the
vertex operators in the theory and for deriving Ward identities like
\sptiwi, but in order to find the exact correlation functions the more
complete form of the vertex operators given in this paper is
necessary.

Another important point has to be discussed.  The expression
\centch\ for the asymptotic form of $I$ is the degree of the
holomorphic map $\gamma(z)$. Clearly it is an integer $q$ independent of
the other insertions in the correlation function
\eqn\iisq{\langle I_f V_1\dots V_n \rangle = q \langle V_1\dots V_n
\rangle,}
and therefore $I_f$ is central.  For example, if $\gamma(z)=z$, $q=1$.
Other values of $q$ can be obtained, \eg\ from $\gamma(z)=z^q$.  Even
though $I_f$ is simply the limit of $I$ as $\phi \rightarrow \infty$,
the expression for the central charge \centch\ differs from the
discussion in section 5 in several crucial respects:
\item{(1)} In section 5 the leading order contribution to the central
charge came, as in \dbort, from a disconnected diagram, while here the
contribution is from a connected diagram.  
\item{(2)} In section 5 $p= \langle I\rangle_0$ was not quantized. This
is consistent with the expectations from the spacetime analysis
described in the introduction that
$p$ is quantized only non-perturbatively.  Here, $q= \langle
I_f\rangle$ is quantized already classically. 
\item{(3)} In the discussion in section 5 connected diagrams also
contribute to $P(g_s)$.  As we saw in \thptd, the leading order
connected contribution is $I_0=1/k$ which is not even quantized. This
is consistent since there is another (disconnected) contribution
at the same order.  In the computation with $I_f$ the entire
contribution comes from the connected diagram.
\item{(4)} The expression \centch\ is non-zero only if the worldsheet
is mapped to an infinite plane in $H_3^+$.  Maps of the worldsheet
to a compact region
in the target space lead to $\langle I_f \rangle =0$.  On the other
hand, we expect that, at least semiclassically, the leading
contribution to the functional integral comes from such maps.  Indeed,
the expressions in section 5 can come from maps to a compact region.

What is then the meaning of $\langle I_f
\rangle=q$?  Clearly, it receives contributions only from worldsheets
at large $\phi$, which have the topology of a plane rather than from
small worldsheets.  Unlike the discussion in section 5, which applies
to short strings propagating in the bulk of $AdS_3$, the discussion of
the central charge here and in \gks\ applies to a worldsheet of a long
string in the region $\phi \rightarrow \infty$.  

Such long strings appeared in \mms\ and a
detailed discussion of them appears in \seiwit.  Recall
that $p$ is the number of fundamental strings creating our background.
Therefore, $q$ coincident long strings act as a domain wall changing
the value of $p$.  On one of its sides we find the original value of
$p$.  On the other side the value of $p$ is screened to $p-q$.  The
effective theory of the long string is a conformal field theory.  The
vertex operators in this theory describe the emission and absorption
of short strings from it.  However, since the worldsheet of the long
string is at large $\phi$, we can safely use the asymptotic form of
the vertex operators.  The worldsheet in this problem is the
worldsheet of the long string which is non-compact.  Furthermore, in
the effective theory of the long string there are no disconnected
worldsheets, as these describe short strings only.  $\langle I_f
\rangle$ measures the number of long strings.  Therefore it is
quantized even at tree level.

This interpretation of the calculation of $\langle I_f \rangle =q$
neatly explains the four puzzles above.

To summarize, the structure emerging from \refs{\gks,\seiwit} 
and this note is the 
following. String theory on $AdS_3$ with a given value of $p$ \defp\ 
contains many different sectors. One sector corresponds to ``short 
strings'', \ie\ maps from the worldsheet of the string to a compact
region in $AdS_3$. This is the sector described in this note. In it,
$p$ determines the fundamental string coupling, $g_s^2\propto 1/p$. The
quantization of $g_s$ is a non-perturbative effect which is not seen
in our classical analysis, as explained in section 1.

Other sectors contain in addition $0<q\le p$ ``long strings'' at infinity. 
These sectors contain two kinds of excitations, fundamental strings 
propagating in the bulk of $AdS_3$ and small ripples on the long strings 
at infinity. 

The analysis of \gks\ plays two roles in string theory on $AdS_3$. 
In the short string sector of the theory, physical observables correspond 
to ``microscopic operators'' in the language of Liouville theory
\mssm; their wavefunctions are exponentially
supported at the boundary $\phi\to\infty$. Thus in order to identify
the observables and study their transformation properties under the
spacetime symmetries, it is enough to analyze the wavefunctions at large
$\phi$; this is done in \gks. In particular, the analysis of \gks\ 
proves that fundamental string states transform in representations
of the infinite spacetime symmetry algebra; it also exhibits the set of
representations that appear in the fundamental string spectrum.

 Correlation functions of these observables (and in particular the
central charge) are not supported at $\phi\to\infty$, and thus 
one needs to analyze the behavior of the theory 
in the bulk of $AdS_3$. 

A second place where the analysis of \gks\ is applicable is in
studying the effective theory of the $q$ long strings located near
the boundary of $AdS_3$. There, the free field description is 
applicable. 
The vertex operators of \gks\ describe small excitations of the 
long strings and the current algebra generators (\eg\ \newll)
can be used to study the transformation properties of these
excitations under the spacetime symmetry algebra. $I_f$ \centch\ 
keeps track of the central charge carried by these long strings.
As one would expect, it is quantized at tree level.

\newsec{Superstrings on $AdS_3\times S^3\times T^4$}

The supersymmetric case adds to the previous discussion a few
new issues, such as the spacetime physics of RR sector fields.
In this section we briefly comment
on the generalization of the construction of the previous sections
to the superstring (see \refs{\gks,\kll} for more details). 

In addition to the bosonic sector of the worldsheet theory we now have
free worldsheet fermions $\psi^A(z)$ which transform in the adjoint
of $SL(2)$ (and a similar structure for the other worldsheet chirality).
The total $SL(2)$ currents $J^A(z)$ now have two contributions: 
a level $k+2$ bosonic current $J^A_B$, and a level $-2$ fermionic
current $J^A_F$; $J^A=J^A_B+J^A_F$. 

It is convenient to follow the discussion of the bosonic case
and use \xevol\ to define
\eqn\psix{\psi(x;z)=2x\psi^3 -\psi^+-x^2\psi^-.}
$\psi(x;z)$ satisfy the OPE algebra\foot{We normalize $\psi^A$
as: $\psi^A(z)\psi^B(w)=\eta^{AB}/(z-w)$.}
\eqn\psixope{\psi(x;z)\psi(y;w)\sim {2(x-y)^2\over z-w}.}
Note that $\psi(x;z)$, like $J(x;z)$ in the bosonic case,
has spacetime scaling dimension $-1$.
The fermionic $SL(2)$ current can be written in terms
of $\psi(x;z)$ as
\eqn\jfxz{J_F(x;z)=\half\psi(x;z)\partial_x\psi(x;z).}
It is easy to check that it satisfies \JJ\ with $k=-2$.
Thus, the total $SL(2)$ current is
\eqn\fulljxz{J(x;z)=J_B(x;z)+
J_F(x;z).}
The worldsheet stress tensor \newstresst\ of the bosonic
degrees of freedom is given by
\eqn\tbws{T_B^{\rm ws}={1\over 2k}\left[ J_B\partial^2_x J_B
-\half(\partial_x J_B)^2\right].}
the fermionic worldsheet stress tensor is 
\eqn\tfws{T^{\rm ws}_F=-\half\psi^A\partial_x\psi_A={1\over
8}\left(\partial_x\psi\partial_z\partial_x\psi 
-\partial_x^2\psi\partial_z\psi-\psi\partial_z\partial_x^2\psi\right),}
and the superconformal generator is (up to an overall constant
which will not play a role below)
\eqn\gws{G^{\rm ws}(z)=-\partial_x\psi\partial_x J_B
+\partial_x^2\psi J_B+\psi\partial_x^2 J_B+\half\psi\partial_x\psi
\partial_x^2\psi.}
As in the bosonic case, it is easy to see that $\partial_x T^{\rm
ws}_F= \partial_x G^{\rm ws}=0$, and that \tfws, \gws\ transform as
tensors of weight $(0,0)$ in spacetime.

The chiral GSO projection implies that the worldsheet theory contains
$(\Delta, \bar\Delta)=(1,0)$ holomorphic operators
\eqn\tamz{\theta_r^{\alpha\mu}=e^{-\half\varphi} S_r^{\alpha\mu},}
where $\varphi$ is the bosonized ghost of the fermionic
string. The operators
$\theta$ have $h=-1/2$ and transform 
in the $(2,2)$ of the $SO(3)\times SO(4)$
symmetry acting on the worldsheet fermions $(\psi^A, \chi^a,
\lambda^i)$ (see \gks\ for notation). $r=\pm\half$ is the $SL(2)$
index; $\alpha=\pm$ is the $SU(2)$ index. $\mu$ denotes a spinor
of $SO(4)$; the other spinor of $SO(4)$ will be denoted by
$\tilde\mu$. There is a second set of holomorphic operators 
$\tilde\theta_r^{\alpha\tilde\mu}$ which will be useful below.

As shown in \gks, $\oint dz \theta_r^{\alpha\mu}$ generate the global
$N=4$ superconformal algebra. To construct the local superconformal
currents it is convenient to define the $x$ dependent spin field
\eqn\thalmu{\theta^{\alpha\mu}(x)=e^{-xJ^-_0}\theta_\half^{\alpha\mu}
e^{xJ^-_0}=\theta_\half^{\alpha\mu}-x\theta_{-\half}^{\alpha\mu}.}
The OPE of $\psi(x;z)$ with $\theta^{\alpha\mu}$ is
\eqn\opepsth{\psi(x;z)\theta^{\alpha\mu}(y;w)\sim {x-y\over (z-w)^\half}
\theta^{\alpha\mu}(x;w)={1\over (z-w)^\half}\left[(x-y)^2\partial_y
+(x-y)\right]\theta^{\alpha\mu}(y;w).}
Furthermore, one can show that 
\eqn\ththab{\theta^{\alpha\mu}(x;z)\theta^{\beta\nu}(y;w)
\sim {\delta^{\mu\nu}e^{-\varphi}\over z-w}
\left[\delta^{\alpha\beta}
\psi(x;w)+(x-y)\sigma_a^{\alpha\beta}\chi^a(w)\right].}

Following the by now familiar pattern, we are looking for
spacetime holomorphic operators involving the worldsheet
current $\theta^{\alpha\mu}$. One can check (using \tbws\ -- \gws)
that the operator  
\eqn\galmu{\Psi^{\alpha\tilde\mu}(x)=-
{1\over\pi k}\int d^2z \tilde\theta^{\alpha\tilde\mu}(x;z)
e^{-\bar\varphi}\bar\psi(\bar x; \bar z)\Phi_1(x,\bar x; z, \bar z)}
is BRST invariant. It
corresponds to a dimension $(\half, 0)$ fermionic operator. 
In fact, the four spacetime fermions $\Psi^{\alpha\tilde\mu}(x)$
are the bottom components of $N=4$ superfields whose top 
components are the $U(1)^4$ currents
\eqn\uonecur{P^i(x)=-{1\over k\pi}\int d^2z
e^{-\varphi-\bar\varphi} \lambda^i(z)\bar\psi(\bar x; \bar
z)\Phi_1(x,\bar x; z, \bar z) ,}
which were constructed in \gks\ and further discussed in \kll. 

At the next level we find the spacetime $N=4$ superconformal
currents 
\eqn\galmust{G^{\alpha\mu}(x)=-{1\over\pi k}
\int d^2z\left(\theta^{\alpha\mu}\partial_x\Phi_1
+2\Phi_1\partial_x\theta^{\alpha\mu}\right)e^{-\bar\varphi}
\bar\psi.}
The form of the stress tensor is an
obvious generalization of \viras, \virasn.
Using the formalism of the previous sections one can show that
the superconformal generators satisfy the $N=4$ superconformal
algebra in spacetime.

Note that just as in the bosonic case studied in
the previous sections, the operators
\galmu\ -- \galmust\ describing holomorphic operators
in the spacetime CFT are ``almost pure gauge.''
The basic observation\foot{In \galmu\ -- \galmust\
we actually need the antiholomorphic version of
the formulae below.} is that the $h=0$ combination
$\exp(-\varphi)\psi(x;z)\Phi_1(x,\bar x; z, \bar z)$
can be written as
\eqn\brstex{
e^{-\varphi}\psi(x;z)\Phi_1(x,\bar x; z, \bar z)=
\{Q_{BRST}, \partial\xi e^{-2\varphi}\Lambda(x, \bar x;
z, \bar z)\},}
where $\xi$ is a field arising in the bosonization of
the bosonic $\beta$, $\gamma$ ghosts in the fermionic
string, and $\Lambda$ is our old friend \ddmm.
The fact that $\Lambda$ is not a good observable
in CFT on $AdS_3$ again implies that despite \brstex, 
$\exp(-\varphi)\psi(x;z)\Phi_1(x,\bar x; z, \bar z)$
is not BRST exact. At the same time its correlation
functions can be computed by a straightforward generalization
of the bosonic analysis. This can be done by working
directly with the $-1$ picture field \brstex, or by  
picture changing it to
the $0$ picture, where it can be written
as a total derivative,
$J_B(x;z)\Phi_1(x,\bar x; z, \bar z)\sim\partial_{\bar z}\Lambda$, 
and one can proceed as in the previous sections.
 
In section 1 we saw from a target space analysis
that RR gauge fields on $AdS_3$ have a mass of the
same order of magnitude $(1/\sqrt k)$ as that of the NS sector
gauge fields. This is manifest in the worldsheet formalism.
The $j$'th partial wave of an NS gauge field obtained by
Kaluza-Klein reduction from ten to three dimensions is described
by the vertex operator
\eqn\Aj{A_j^i=e^{-\varphi}\psi(x;z)e^{-\bar\varphi}\bar\lambda^i(
\bar z)\Phi_{j+1}(x,\bar x) V_j(z,\bar z),}
where $V_j$ is the wavefunction on $S^3$ (see \refs{\gks,\kll}
for details). $A_j$ transforms in the spin $(j,j)$ representation
of $SU(2)_L\times SU(2)_R$ and has spacetime scaling
dimension
\eqn\scdimAR{(h,\bar h)=(j, j+1).}
The corresponding RR gauge field is described by the
vertex operator
\eqn\Rj{R_j^{\tilde\mu\bar\mu}=e^{-{\varphi\over2}-{\bar\varphi
\over 2}}\tilde\theta^{\alpha\tilde\mu}\left(
\bar\theta^{\bar\alpha\bar\mu}\partial_{\bar x}\Phi_{j+\half}
+2\Phi_{j+\half}\partial_{\bar x}\bar\theta^{\bar\alpha\bar\mu}
\right)V_{j-\half},}
where $\tilde\theta$, $\bar\theta$ and $V$ are coupled into a 
representation with total spin $(j,j)$ under $SU(2)_L\times SU(2)_R$
(as \Aj). The spacetime scaling dimension of \Rj\ is again given by
\scdimAR. In fact, the operators $(A_j, R_j)$ together with their
antiholomorphic counterparts form multiplets of the U-duality
group. These multiplets were studied in \kll.

In section 1 we also saw that string theory on
$AdS_3\times S^3\times T^4$ is expected to have in addition
to the NS $U(1)_L^4\times U(1)_R^4$ current algebra described
on the worldsheet by \uonecur\ and its right moving analog,
a $U(1)_L^4\times U(1)_R^4$ current algebra associated with
RR gauge fields (see the discussion after \rrterms). One would
expect this current algebra to correspond to RR vertex operators
on the worldsheet. We have been unable to find vertex operators
with the right properties. It is possible that we have not looked
carefully enough.  However, since the spacetime CFT is two dimensional
and there are no perturbative string states which are charged under
this RR current algebra, it forms a decoupled sector of the theory.  
Therefore, it is not clear whether such vertex operators should be 
found in our worldsheet approach.

\newsec{Summary}

In this paper we studied string theory in an $AdS_3$ background.  Most of
the discussion did not depend on many of the details of how this
background is generated.  On general grounds, the three dimensional
low energy effective
Lagrangian is \stringlag\
\eqn\stringlagc{e^{-2\phi} \sqrt{-g} \big( {4 \over k} + R + 4(\nabla
\phi)^2 - {1 \over 12} H_{\mu\nu\rho}^2 \big).}
The equations of motion show that $H$ is nonzero and the zero mode of
$\phi$ is an arbitrary integration constant.  The nonzero $H$ can be
interpreted as arising from a source of strength $p={4\pi \over
\hbar\sqrt{k}}$ at infinity.  Classically and to all orders in
perturbation theory $p$ is an arbitrary real number.  When there is
a worldsheet current algebra $\widehat G$ at level $k_G$, there are
gauge fields in spacetime and the Lagrangian \stringlagc\ is modified
to \stringlaggau.  As a result, $\widehat G$ current algebra at level
$k_G^{(st)}=k_Gp$ appears in the spectrum.  Hence,
nonperturbatively $p$ must be quantized.

We then studied various properties of the worldsheet CFT on $AdS_3$.
Following \zamfat\ we parameterized the operators in terms of their
$SL(2)$ quantum number $h$ and a continuous complex parameter $x$,
which is interpreted as a coordinate on the boundary of the space, on 
which the spacetime CFT ``lives.''  The operator
\ooyy
\eqn\phionec{\Phi_1={1 \over \pi} \left({1 \over (\gamma-x)(\bar\gamma
- \bar x) e^\phi + e^{-\phi}}\right)^2,}
which has worldsheet conformal dimensions $(0,0)$ and spacetime dimensions
$(1,1)$ plays an important role in the theory.  
Using this operator, the worldsheet $SL(2)$ current $J(x;z)$
and $\widehat G$ worldsheet currents $k^a$, we
constructed the vertex operators of the spacetime current algebra
\genkb, the spacetime identity operator \newp, and spacetime Virasoro 
algebra \viras\
\eqn\genkbc{\eqalign{
&K^a(x)= -{1\over k}\int d^2 zk^a(z)
\bar J(\bar x; \bar z)\Phi_1(x, \bar x; z, \bar z) \cr
&I={1\over k^2} \int d^2z J(x;z)\bar J(\bar x;
\bar z)\Phi_1(x,\bar x; z,\bar z) \cr
&T(x)={1\over2k}\int d^2z \left(\partial_xJ\partial_x\Phi_1+
2\partial_x^2J\Phi_1 \right) \bar J(\bar x;\bar z) .\cr}}
Assuming \opesems
\eqn\opesemsc{\lim_{z_1\rightarrow z_2} 
\Phi_1(x_1,\bar x_1; z_1, \bar z_1) \Phi_h(x_2,\bar x_2; z_2, \bar
z_2) = \delta^2(x_1-x_2) \Phi_h(x_2,\bar x_2; z_2, \bar z_2),} 
which can be proven in the leading order in the large $k$ semiclassical
expansion, we showed that \genkbc\ satisfy the correct
operator product expansion in spacetime 
\eqn\stkmvirwi{\eqalign{
K^a(x)K^b(y)\sim &{\half k_G^{(st)}\delta^{ab}\over (x-y)^2}
+{f^{abc} K^c(y)\over x-y}\cr
K^a(x)W_h(y, \bar y)\sim &          
{t^a(R)W_h(y, \bar y)\over x-y}\cr 
T(x)T(y)\sim&{\half c^{(st)}\over (x-y)^4}
+{2T(y)\over (x-y)^2}+{\partial_y T\over x-y}\cr
T(x)W_h(y, \bar y)\sim&          
{hW_h(y, \bar y)\over (x-y)^2}+{\partial_y W_h\over x-y}.\cr
}}
The fact that $I$ in
\genkbc\ behaves like a spacetime identity operator follows from
another assumption which we did not prove, \assumei.

As we stressed, since the first quantized string description is in
Landau gauge, we do not have the freedom to gauge the operators
\genkbc\ to the boundary.  Therefore the corresponding vertex
operators have nontrivial profiles in the bulk of $AdS_3$.

Extending these techniques to the superstring we found the expected
$N=4$ spacetime superconformal algebra.

As $\phi \rightarrow \infty$, one recovers the results of \gks.  The
discussion of \gks\ can be used for two distinct purposes in string
theory on $AdS_3$. In the ``short string'' sector it is useful for
enumerating the observables and studying their transformation
properties under the spacetime symmetry algebra.  It can also be used
to study the physics of long strings in $AdS_3$.  One important
difference between the physics of short strings and long strings is
the value of the central charge.  For short strings the central charge
arises (in leading order), as in \dbort, from disconnected diagrams.
As expected from the spacetime analysis, it is not quantized in
perturbation theory.  For long strings the central charge arises as in
\gks\ from a cylindrical connected diagram.  It is quantized at the
leading order.

\bigskip
\centerline{\bf Acknowledgements}
We thank O. Aharony, S. Elitzur, F. Larsen, J. Maldacena, G. Moore,
E. Rabinovici, A. Schwimmer, A. Strominger, E. Witten and
A.B. Zamolodchikov for discussions and J. Maldacena for helpful
suggestions on the manuscript.  We also thank the organizers of the
String Workshop at the Institute for Advanced Studies at the Hebrew
University for hospitality and for creating a stimulating environment
during the completion of this work.  D.K. was supported in part by DOE
grant \#DE-FG02-90ER40560 and N.S. by DOE grant \#DE-FG02-90ER40542.

\listrefs
\bye